\documentclass{article}

\usepackage{microtype}
\usepackage{graphicx}
\usepackage{subfigure}
\usepackage{booktabs} 

\usepackage{hyperref}
\usepackage{colortbl}

\usepackage[accepted]{icml2024}

\usepackage{amsmath}
\usepackage{amssymb}
\usepackage{mathtools}
\usepackage{amsthm}
\usepackage{array}
\usepackage{multirow}
\usepackage[all=normal, mathspacing=tight,mathdisplays=normal, floats=tight, paragraphs=tight, lists=normal]{savetrees}

\renewcommand{\b}{\boldsymbol}

\definecolor{FadedBlack}{rgb}{0.2, 0.2, 0.2} 

\colorlet{FadedBlack}{black!50} 

\usepackage[capitalize,noabbrev]{cleveref}

\theoremstyle{plain}

\theoremstyle{definition}

\theoremstyle{remark}

\usepackage[textsize=tiny]{todonotes}
\usepackage{siunitx}

\icmltitlerunning{DiscDiff: Latent Diffusion Model for DNA Sequence Generation}

\begin{document}

\twocolumn[
    \icmltitle{DiscDiff: Latent Diffusion Model for DNA Sequence Generation}



    \icmlsetsymbol{equal}{*}

    \begin{icmlauthorlist}
        \icmlauthor{Zehui Li}{yyy}
        \icmlauthor{Yuhao Ni}{sch}
        \icmlauthor{William A V Beardall}{yyy}
        \icmlauthor{Guoxuan Xia}{sch}
        \icmlauthor{Akashaditya Das}{yyy}
        \icmlauthor{Tim August B. Huygelen}{zzz}
        \icmlauthor{Guy-Bart Stan}{yyy}
        \icmlauthor{Yiren Zhao}{sch}
    \end{icmlauthorlist}

    \icmlaffiliation{yyy}{Department of Bioengineering, Imperial College London}
    \icmlaffiliation{sch}{Department of Electrical and Electronic Engineering, Imperial College London}
    \icmlaffiliation{zzz}{University College London}
    \icmlcorrespondingauthor{Zehui Li}{zl6222@ic.ac.uk}
    \icmlcorrespondingauthor{Guy-Bart Stan}{g.stan@imperial.ac.uk}
    \icmlcorrespondingauthor{Yiren Zhao}{a.zhao@imperial.ac.uk}

    \icmlkeywords{Machine Learning, ICML}

    \vskip 0.3in
]



\printAffiliationsAndNotice{} 

\begin{abstract}
    This paper introduces a novel framework for DNA sequence generation, comprising two key components: DiscDiff, a Latent Diffusion Model (LDM) tailored for generating discrete DNA sequences, and Absorb-Escape, a post-training algorithm designed to refine these sequences. 
    Absorb-Escape enhances the realism of the generated sequences by correcting `round errors' inherent in the conversion process between latent and input spaces. Our approach not only sets new standards in DNA sequence generation but also demonstrates superior performance over existing diffusion models, in generating both short and long DNA sequences. Additionally, we introduce EPD-GenDNA, the first comprehensive, multi-species dataset for DNA generation, encompassing 160,000 unique sequences from 15 species. We hope this study will advance the generative modelling of DNA, with potential implications for gene therapy and protein production.
\end{abstract}

\section{Introduction}
\label{sec:introduction}

Diffusion Models (DMs) are emerging as a promising tool in the field of protein design -- models using structural~\citep{watson2023novo}
and amino acid sequence level data have produced state-of-the-art results ~\citep{alamdari2023protein}. On the other hand, the application of DMs for DNA generation remains limited. Well-designed synthetic DNA sequences are vital for enhancing gene therapies tasks~\citep{liu2022cancer}, such as the control of gene expressions in complex genetic circuits~\citep{zrimec2022controlling}
and the production of novel proteins ~\citep{gupta2019feedback}.
However, research in applying deep learning to these important tasks is heavily limited by the absence of extensive datasets for DNA generation, well-defined evaluation metrics, and robust baseline models.

Prior work on DNA generation has used Generative Adversarial Networks (GANs)~\citep{gan}
for the generation of synthetic DNA sequences~\citep{killoran2017generating,wang2020synthetic,zrimec2022controlling}. However, it has been shown that the generated samples lack diversity~\citep{dhariwal2021diffusion} and suffer from mode collapse at
training time~\citep{metz2016unrolled}. To address these issues, Diffusion Models (DMs) have been proposed, which iteratively refine sample quality. Despite their effectiveness, the discrete nature of DNA sequences poses a unique challenge for standard DMs, traditionally designed for vision data. This has led to the development of specialized variants like the Dirichlet Diffusion Score Model (DDSM)~\citep{avdeyev2023dirichlet} and EvoDiff~\cite{alamdari2023protein}, tailored for discrete biological sequence generation. These models employ diffusion-like processes to modify discrete data, although they are computationally intensive, as discussed by~\citep{salimans2017pixelcnn++}. 

In an attempt to mitigate computational demands, Latent Diffusion Models (LDMs) have been introduced, utilizing a compressed and regularized latent space to reduce complexity. However, current LDMs for discrete data~\citep{dieleman2022continuous, li2022diffusion, han2022ssd} are mainly domain-specific and not directly applicable to DNA sequences. A significant challenge in adapting LDMs for discrete data generation lies in mitigating `rounding errors' during conversions between the latent and input spaces. Such errors can lead to inaccuracy, especially in long sequences. This problem has been evident in text generation, where LDMs often produce less fluent sentences compared to autoregressive models~\citep{zhang2023planner}.

\begin{figure}[ht]
      \vskip 0.2in
      \begin{center}
            \centerline{\includegraphics[width=\columnwidth]{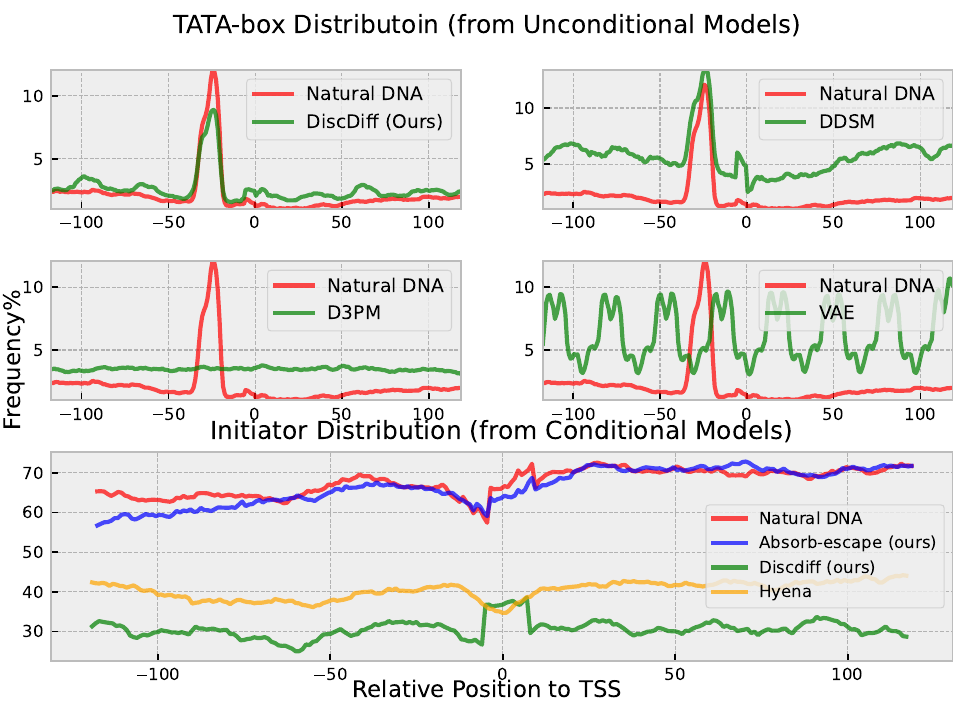}}
            \caption{A comparison of Motif frequency distributions. The graphs contrast the occurrences of TATA-Box and Initiator motifs at each position in a set of samples from natural DNA against those generated by various models. A close match in frequency distributions suggests a higher realism and better performance for the generated DNA sequences. \textbf{DiscDiff and Absorb-Escape outperform existing models qualitatively by a significant margin.}}
            \label{fig:front-page}
      \end{center}
      \vskip -0.2in

\end{figure}

This study introduces a novel learning framework for DNA sequence generation. It consists of: 1) DiscDiff - an LDM for generating discrete sequences and 2) Absorb-Escape - a post-training algorithm to refine sequences generated by the LDM. DiscDiff contains a new network architecture for mapping discrete sequences to continuous latent spaces and vice versa. Absorb-Escape, on the other hand, is a novel inference framework that combines the strengths of LDMs and autoregressive models. Absorb-Escape scans through the generated sequences, and actively corrects these `rounding errors', thus producing more realistic samples.  As shown in~\Cref{fig:front-page}, while DiscDiff by itself has achieved state-of-the-art results in DNA sequence generation, Absorb-Escape further improves the quality of the generated DNA sequences. In addition, to fully evaluate the potential of these techniques, we propose the first large-scale cross-species dataset for DNA generation, with defined metrics and tasks to benchmark generative models on DNA synthesis. Our contributions are as follows:

\begin{table}[t]
\centering
\caption{A comparison of DNA generation datasets, our EPD-GenDNA is significantly larger in size and contains more species.}
\resizebox{\linewidth}{!}{
\tiny 
\setlength{\tabcolsep}{3pt} 
\begin{tabular}{@{}llll@{}} 
\toprule
Dataset & \# DNA & Species & Seq. Len. \\ \midrule
EPD-GenDNA (Ours) & 160k & 15 & 2048\&256 \\
DDSM~\citep{avdeyev2023dirichlet} & 100k & Human & 1024 \\
ExpGAN~\citep{zrimec2022controlling} & 4238 & Yeast & 1000 \\
EnhancerDesign~\citep{taskiran2023cell} & 7770 & Human & 500 \\
\bottomrule
\end{tabular}
}
\label{table:dataset}
\end{table}

\begin{table}[t]
\centering
\caption{A comparison of DNA generation algorithms, our DiscDiff is the first LDM approach that is tested on various species.}
\resizebox{\linewidth}{!}{
\tiny 
\setlength{\tabcolsep}{3pt} 
\begin{tabular}{@{}llcl@{}} 
\toprule
Algorithm & \ Type & Regulatory/Protein Encoding Regions & MultiSpecies \\ \midrule
DiscDiff (Ours) & LDMs & Both & $\checkmark$ \\
DDSM~\citep{avdeyev2023dirichlet} & DMs & Both & $\times$ \\
ExpGAN~\citep{zrimec2022controlling} & GANs & Regulatory & $\times$ \\
EnhancerDesign~\citep{taskiran2023cell} & GANs & Regulatory & $\times$ \\
FeedbackGAN~\citep{gupta2019feedback} & GANs & Protein Encoding & $\times$ \\
\bottomrule
\end{tabular}
}
\label{table:algorithm-intro}
\end{table}

\begin{itemize}
    \item We introduce EPD-GenDNA, a dataset built on EPD~\citep{meylan2020epd}. As shown in \Cref{table:dataset}, it is the first comprehensive, multi-species, multi-cell type dataset for DNA generation. The dataset encompasses 160,000 unique DNA sequences, derived from 15 species, covering 2,713 cell types, and is based on 300 million wet lab experiments. 

    \item We propose \textit{DiscDiff}, to our best knowledge, the first application of Latent Diffusion Models (LDMs) to the task of DNA generation. DiscDiff surpasses existing state-of-the-art diffusion models in short and long DNA generation by 7.6\% and 1.9\%, respectively, as measured by the motif distribution.

    \item We propose \textit{Absorb-Escape}, a generalisable post-training algorithm for refining the quality of generated discrete sequences. We show that Absorb-Escape further increases the performance of DiscDiff by 4\% in long DNA generation. In addition, Absorb-Escape allows control over the property of generated samples.

    \item  Lastly, we demonstrate a practical application of DNA generative modelling. We train a conditional generative model capable of realistically simulating regulatory elements and genes across 15 species.
\end{itemize}

\section{Related Work}

\paragraph{Diffusion Models (DMs).}

Initially proposed by~\citep{sohl2015deep}, two parallel perspectives of diffusion models have been developed. In the probalistic denoising view~\citep{ho2020denoising},
a forward diffusion process gradually adds noise to the input data $x_0$, and a reverse diffusion process gradually removes the noise from the perturbed data $x_t$.

The reverse denoising (generative) process is parameterized with a trainable network $\mu_\theta(x_t,t)$.
$\mu_\theta(x_t,t)$ is either a U-Net~\citep{ronneberger2015u} or a Transformer~\citep{gupta2023photorealistic} with free parameters, and is
trained via evidence lower bound (ELBO) or denoising score matching loss.

An alternative perspective involves modelling diffusion as stochastic differential equations (SDE) and their respective inverse processes.
~\citep{song2020score}. This formulation allows the use of
ODE solvers to compute the reverse process efficiently~\citep{song2020denoising,lu2022dpm,zhang2023fast,Xia2023ScoreNF}.
In this work, we will follow the denoising view of the diffusion model, and use a reparameterising trick found in Latent Diffusion Models~\citep{rombach2022high}. In this notation, the training process can also be interpreted as training a noise predictor $\mathbf{\varepsilon_\theta}$:
\begin{equation}
    \label{eq:training}
    \min_{\theta} \mathbb{E}_{x,t \sim U[1,T],\varepsilon \sim \mathcal{N}(0,1)} \left[  \|\varepsilon -\mathbf{\varepsilon_\theta}(x_t,t)  \|^2_2 \right].
\end{equation}

\paragraph{Latent Diffusion Models (LDMs).}
LDMs convert discrete inputs into a continuous latent space. This technique is widely employed in language modelling, where text is transformed into a continuous format using tools like word2vec or neural network-based embeddings~\citep{dieleman2022continuous, li2022diffusion, han2022ssd}. Diffusion-like processes can be applied to discrete data 
\cite{avdeyev2023dirichlet}, but LDMs are preferred for their computational efficiency - this is due to the  compaction and smoothness of the latent space~\citep{rombach2022high}. Current state-of-the-art models are usually domain-specific and rely on pre-trained language models to form latent embeddings ~\citep{lovelace2022latent, zhang2023planner}. Additionally, discrete generation using diffusion models often encounters rounding errors when mapping from the latent space to text tokens \citep{zhang2023planner, li2022diffusion, lin2022genie}. Similar issues arise in the generation of DNA sequences. Specifically, the accumulation of rounding errors can result in the production of invalid DNA sequences, this effect is more prominent in the generation of longer sequences.

In contrast, autoregressive models learn the conditional probability of the next token based on previous ones. This generally results in more locally coherent samples. Such models have been successfully used in text generation~\citep{radford2019language,brown2020language,touvron2023llama} and genomic sequence modeling~\citep{lal2023reglm, ji2021dnabert,dalla2023nucleotide,nguyen2023hyenadna}.
However, autoregressive models can accumulate errors over time, leading to repetitive and overly confident outputs~\citep{xu2022learning}. They generate samples with lower diversity than diffusion models~\citep{lovelace2022latent}, which is detrimental to genome sequence generation, where diveristy is a key metric.

While there are works aiming to combine the advantages of diffusion and autoregressive models~\citep{zhang2023planner, xu2022learning}, they rely on existing
autoregressive language models, which have the same time and space complexity as the original autoregressive models. In this work, we propose a novel algorithm, the \textit{Absorb-Escape} algorithm, which can be applied to any diffusion model to improve the local coherency of generated samples.

\paragraph{Deep Generative Models for DNA Sequences}
The primary goal of DNA generation is to produce \textit{realistic} regulatory elements or genes. Regulatory elements~\citep{eraslan2019deep} are specific DNA sequences that control gene expression, whereas genes encode proteins. Generative adversarial networks (GANs) have been used for DNA sequence creation. This includes the creation of protein-encoding region~\citep{wang2020synthetic}, enhancer~\citep{taskiran2023cell} and promoter sequences~\citep{zrimec2022controlling}.

While diffusion models for DNA generation are less explored, one recent work Dirichlet Diffusion Score Model (DDSM) \citep{avdeyev2023dirichlet} uses the Dirichlet distribution in the discrete diffusion process. DDSM also compares its performance against other discrete DMs, such as BitDiffusion~\citep{chen2022analog} and D3PM~\citep{austin2021structured} on the human DNA generation task. However, DDSM primarily focuses on conditional generation, where a real-valued vector of the same length as the input sequence is used for conditioning. This approach simplifies the DNA generation problem, due to the large amount of conditional information available for each sample. 

Large autoregressive models have also been adopted for learning DNA sequence representations~\citep{ji2021dnabert, dalla2023nucleotide, nguyen2023hyenadna}.
This includes transformer-based models~\citep{ji2021dnabert, dalla2023nucleotide} and state-space models
~\cite{nguyen2023hyenadna,lal2023reglm,gu2023mamba}. Notably, state-space models with fewer parameters can outperform transformer-based architectures in both genomics classification and regression tasks~\citep{gu2023mamba}.

The development of generative algorithms for DNA sequences has significant challenges, primarily due to the scarcity of well-formatted data and effective evaluation metrics. While existing datasets support regulatory element~\citep{grevsova2023genomic}, species classification~\citep{mock2022taxonomic}, and genomic assay prediction~\citep{kelley2018sequential,avsec2021effective}, prior DNA generation efforts have been limited by small, single-species datasets~\citep{zrimec2022controlling,wang2020synthetic,avdeyev2023dirichlet,killoran2017generating}. Although motif distribution plots have been used for evaluating the property of the generated DNA sequences, there is a lack of metrics available to quantitatively measure the quality of generated sequences.

To address this gap, our study extends the current analysis meta to include a comparative evaluation between autoregressive and diffusion models for DNA sequence generation. We explore the unconditional and conditional generation of long and short DNA sequences. We assess the effectiveness of the generation process using models trained on a comprehensive, large-scale dataset and analyse their performance using quantitative metrics. This work provides a thorough method for the comparison of generative models in the DNA generation space and advances the field.

\section{Generation Task with EPD-GenDNA}
\label{sec:task}

We present EPD-GenDNA, a novel multi-species DNA sequence dataset, originating from the Eukaryotic Promoter Database (EPDnew)~\citep{meylan2020epd}. This extensive dataset includes manually curated and authenticated DNA sequences, along with comprehensive annotations specifying species and cell types for each sequence. Incorrect annotations were filtered out, with detailed preprocessing steps described in \Cref{appendix:dataset}.

\paragraph{Dataset Format}
\begin{figure*}[ht]
    \vskip 0.2in
    \begin{center}
    \centerline{\includegraphics[width=\linewidth]{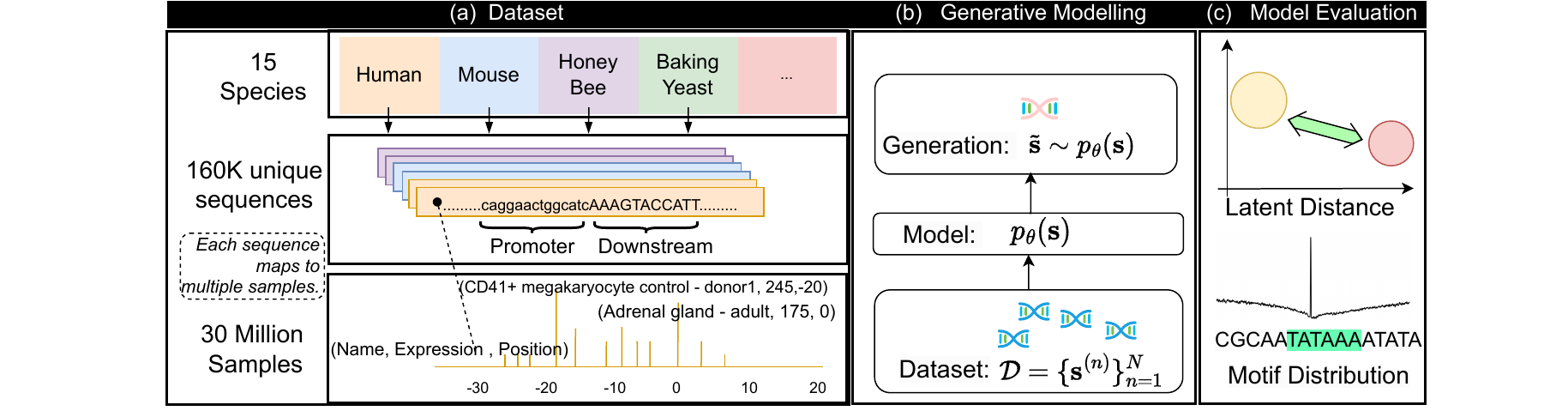}}
    \vspace{-10pt}
    \caption{Generation Task with EPD-GenDNA. \textbf{(a) Dataset:} The EPD-GenDNA dataset includes 160K unique sequences from 15 species and 30 million samples with associated metadata.
\textbf{(b) Generative Modelling:} A probabilistic model $p_{\theta}(s)$ is trained to generate new DNA sequences. 
\textbf{(c) Model Evaluation:} Generated sequences are evaluated through measuring latent distances and analyzing motif distributions.}
        \label{pic:epd-dataset}
    \end{center}
    \vspace{-20pt}
\end{figure*}

As shown in \Cref{pic:epd-dataset}, EPD-GenDNA includes 160K unique (promoter, gene) pairs across 15 species, covering a significant portion of known mammalian promoters~\citep{perier1998eukaryotic}. The dataset spans 30 million samples, each associated with metadata like sample names, cell types, and expression levels, detailed in \Cref{table:epdnew} in \Cref{appendix:dataset}. We present two versions of the dataset with different sequence lengths.
1) \textbf{EPD-GenDNA-small}: 256-base window centered on a gene's transcription start site, split into upstream promoter and downstream segment.
2) \textbf{EPD-GenDNA-large}: 2048-base window centered on a gene's transcription start site, for broader genetic regions.

While the small version is suitable for single regulatory elements and small gene generation, the large version is more suitable for the generation of multiple regulatory elements and large genes. Let \( \mathbb{N}_4 = \{1, 2, 3, 4\} \) where each element represents one of the four nucleotides in a DNA sequence, specifically, 
adenine (A), thymine (T), guanine (G), and cytosine (C).
Then, a DNA sequence of length $L$ can be represented as \( \b s \in \mathbb{N}^L_4 \). In the above datasets, $L$ is 256 and 2048, respectively.

\paragraph{The Unconditional Generation Task}
We work with a dataset of real-world DNA sequences \(\mathcal{D} = \{\b s^{(n)}\}_{n=1}^N\) collected from some distribution $p(\b s)$, where each sequence \(\b s^{(n)} \in \mathbb{N}^L_4\) represents a chain of nucleotides. \(L = 256\) for EPD-GenDNA-small and \(L = 2048\) for EPD-GenDNA-large.
The objective is to develop a generative model \(p_{\b \theta}(\b s)\) of the data distribution $p(\b s)$ from which we can sample novel sequences \(\tilde{\b s}\sim p_{\b\theta}(\b s)\).
These sequences should be structured arrangements of A, T, G, and C, reflecting the complex patterns found in actual DNA.

\paragraph{The Conditional Generation Task} The conditional generation task involves generating DNA sequences given specific conditions or attributes. Here, the dataset of DNA sequences \(\mathcal{D} = \{\b s^{(n)}, c^{(n)}\}_{n=1}^N\) is sampled from the joint distribution $p(\b s, c)$, where \(c\) represents the condition associated with each sequence. The conditions could include species type, expression level, or cell types, adding a layer of specificity to the generation process.
The objective in this task is to develop a model \(p_{\b\theta}(\b s|c)\) that generates new DNA sequences \(\tilde{\b s}\) given condition \(c\). In this paper, we focus on conditioning DNA generation with the species of the organism.

\paragraph{Evaluation Metrics}

For the evaluation of generated DNA sequences, the following metrics are used:
\begin{enumerate}
    \item \textbf{Motif Distribution Correlation (\(\text{Cor}_{\text{M}}\))}: 
    The motifs are small sequences carrying biological significance. And motif distribution is the frequency distribution of the motif along the length of a set of DNA sequences. As shown in \Cref{fig:front-page}, motif distributions of the generated DNA sequences should be similar to the real DNA sequences. More precisely,  \(\text{Cor}_{\text{M}}\) measures the correlation between generated and natural DNA sequences for specific motifs frequency, where $M \in \{\text{TATA-box},\text{GC-box}, \text{Initiator},\text{CCAAT-box}\}$.
    The correlation is defined as the Pearson correlation between the motif distributions of generated and natural DNA sets.
    
    \item \textbf{Diversity}: Defined as \(\prod_{n=10}^{12} \frac{|f_{\text{count}}(n,\mathcal{D})|}{f_{\text{count}}(n,\mathcal{D})}\), where \(f_{\text{count}}\) counts the number of n-grams in a set of generated DNA sequences \(\mathcal{D}\), and \(|f_{\text{count}}(n,\mathcal{D})|\) is the unique number of n-grams. This metric assesses the variety within the generated DNA sequences.
    
    \item \textbf{S-FID (Sei Fréchet Inception Distance)}: Adapted from the Fréchet Inception Distance used in image generation~\citep{heusel2017gans}, S-FID metric measures the distance between the distributions of generated and natural DNA sequences in the latent space. It uses the encoder of a pre-trained genomic neural network, Sei~\citep{zhou2015predicting}, to map the DNA sequence into the latent space. 
\end{enumerate}

\section{Method}

\begin{figure*}[!t]
    \begin{center}
    \centerline{\includegraphics[width=\textwidth]{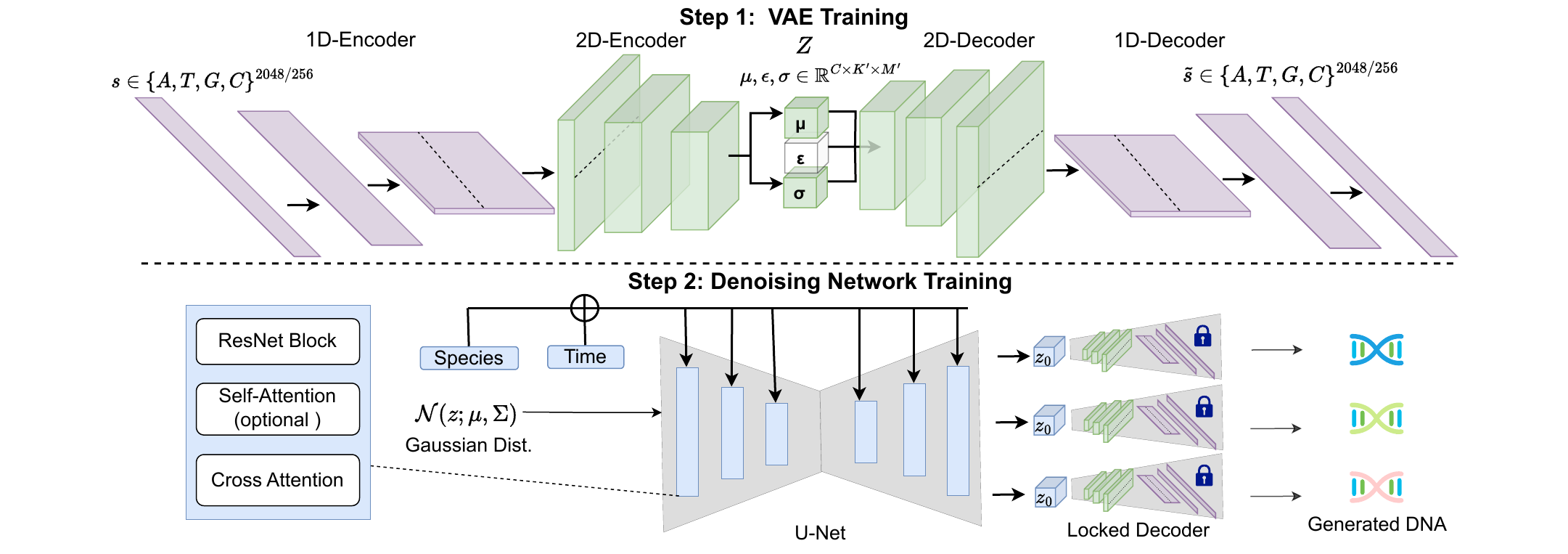}}

\caption{DiscDiff Model: A two-step process for DNA sequence generation. \textbf{Step 1: VAE Training}: 
A sequence $s \in \{A, T, G, C\}^{2048/256}$ is encoded via a 1D-Encoder to a 2D-Encoder. The latent space representation $Z$ with parameters $\mu, \epsilon, \sigma$ is then decoded back to $\tilde{s}$ through a 2D-Decoder and 1D-Decoder.
\textbf{Step 2: Denoising Network Training}: 
The latent representation $Z$ is processed through a denoising network comprising a ResNet Block, optional Self-Attention, and Cross Attention, with species and time information. The network outputs a Gaussian distribution $N(z; \mu, \Sigma)$. A U-Net architecture takes this distribution to produce various $z_0$ representations, which a Locked Decoder (fronzen parameters) used to generate the final DNA sequences.}
        \label{pic:discdiff}
    \end{center}
\end{figure*}

\subsection{The DiscDiff Model}

We introduce DiscDiff, a novel, and to our knowledge the first Latent Discrete Diffusion model for DNA generation tasks.
As shown in \Cref{pic:discdiff}, this model is structured into two main components: a Variational-Auto-Encoder (VAE) and a denoising model. The VAE consists of an encoder $\mathbf{E}: \b s \mapsto \b z$, which maps discrete input sequence $\b s$ to a continuous latent variable $\b z$, and a decoder $\mathbf{D}: \b z \mapsto \tilde{\b s}$, which reverts $\b z$ back to $\tilde{s}$ in the discrete space.
The denoising model $\mathbf{\b\varepsilon_{\b\theta}}(\b z_t,t)$ is trained to predict the added noise $\b\varepsilon$ in the latent space.

\subsubsection{VAE Architecture}

The choice of VAE architecture in LDMs is critical and often domain-specific. For example, Stable Diffusion~\citep{rombach2022high} uses
a hybrid architecture that combines convolutional neural networks (CNNs) and 2d-attention for image generation,
while PLANNER~\citep{zhang2023planner} and LD4LG~\citep{lovelace2022latent} use transformer-based architectures for text generation.

Three types of encoders have been widely used in genomic sequence modelling:
CNNs with pooling layers~\citep{alipanahi2015predicting,kelley2018sequential,kelley2020cross}, CNNs with self-attention~\citep{avsec2021effective}, and 1D-Swin~\citep{li2023genomic}.
In this work, we design three autoencoders based on these achiectecture variants. Namely, we have CNN-VAE, CNN-Attn-VAE, and 1D-Swin-VAE.
Furthermore, we find that mapping the input data to a higher dimension space can help to learn a better denoising network, generating more realistic DNA sequences as shown in the ablation study of \Cref{sec:ablation}.
We hereby propose to use a two-stage VAE architecture as shown in \Cref{pic:discdiff}.

The first stage encoder $\mathbf{E_{\b\phi}}_1: \mathbb{N}^L_4 \rightarrow \mathbb{R}^{K \times M}
$ maps $\b s \in \mathbb{N}^L_4$ to a 2D latent space $\b z_1 \in \mathbb{R}^{K \times M}$, where $K$ is the number of channels and $M$ is the length of the latent representation.
The second stage encoder $\mathbf{E_{\b\phi}}_2 : \mathbb{R}^{1 \times K \times M} \rightarrow \mathbb{R}^{ C \times K' \times M'}$ first adds a dummy dimension to $\b z_1$ such that $\b z_1 \in \mathbb{R}^{1 \times K \times M}$ and then maps it to 3d latent space $\b z \in \mathbb{R}^{ C \times K' \times M'}$,
where $C$ is the number of channels, $K'$ and $M'$ are the reduced dimensions of $K$ and $M$ respectively.
The decoder in the first and second stage are $\mathbf{D_{\b\theta}}_1$ and $\mathbf{D_{\b\theta}}_2$ respectively. Which are symmetric to the encoders. Overall,
we have $\b z = \mathbf{E_{\b\phi}}(\b s) = \mathbf{E_{\b\phi}}_2(\mathbf{E_{\b\phi}}_1(\b s))$, and the reconstruction is $\tilde{\b s} = \mathbf{D_{\b\theta}}(\b z) = \mathbf{D_{\b\theta}}_1(\mathbf{D_{\b\theta}}_2(\b z))$.

\subsubsection{VAE Loss} 

When training the VAE, we propose to use Cross Entropy (CE) as reconstruction loss. The loss function is given by:

\begin{align*}
    \mathbf{L}_{\theta,\phi} = \underbrace{
        \mathbb{E}_{p(\b s)}\left[\mathbb{E}_{q_{\b \phi}(\b z|\b s)}\left[
            -\sum_{l=1}^{L}\sum_{i=1}^{4}\delta_{is_l} \log p_{\b\theta}(\b s_l|\b z)
    \right]\right]}_{\text{Reconstruction Loss}} + \\
    \beta\cdot \underbrace{\mathbb{E}_{p(\b s)}\left[
            \mathrm{KL}(q_{\b\phi}(\b z|\b s) \,||\, \mathcal{N}(\b z; \b \mu, \b \Sigma))
            \right]}_{\text{KL Divergence}}
\end{align*}

where $\delta_{ij}$ is the Kronecker delta, \( p_{\b\theta}(\b s|\b z) \) is the probabilistic decoder output from $\mathbf{D_{\b\theta}}$; \( q_{\b\phi}(\b z|\b s) \) is the probabilistic output from encoder $\mathbf{E_{\b\phi}}$ that represents the approximate posterior of the latent variable \(\b z \) given the input \( \b s \); $\mathcal{N}(\b z;\b \mu, \b\Sigma)$ is the prior on \( \b z \). Here we use a simple isotropic. $\beta$ is a mixing hyperparameter.

\subsubsection{Denoising Network Training}
Once $\mathbf{D_{\b\theta}}$ and $\mathbf{E_{\b\phi}}$ are trained in the first step,
we train a noise prediction $\b\varepsilon_{\b\theta}$ in the latent space $\b z = \mathbf{E_{\b\phi}}(\b s)$ with
\Cref{eq:denoising-loss}. 
The details about the U-Net configuration can be found in \Cref{app:unet-config}.
\begin{equation}
    \mathbb{E}_{\b z,t \sim U[1,T],\b\varepsilon \sim \mathcal{N}(\b 0,\b I)} \left[  \|\b\varepsilon -\mathbf{\b\varepsilon_{\b\theta}}(\b z_t,t)  \|^2_2 \right]
    \label{eq:denoising-loss}
\end{equation}

\subsection{Absorb-Escape}
\label{sec:Absorb-Escape}
\begin{figure}[t]
    \vskip 0.2in
    \begin{center}
        \centerline{\includegraphics[width=\columnwidth]{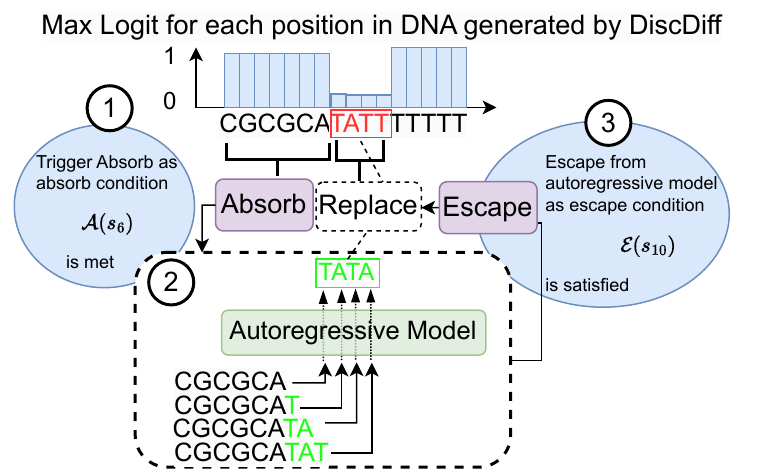}}
        \caption{The Absorb-Escape Algorithm: Enhancing DNA Sequence Prediction. While diffusion models (DMs) effectively capture broad DNA sequence features, they can err at the single nucleotide level. The Absorb-Escape algorithm corrects such errors by identifying and modifying low probability nucleotides, like changing `TATT' to `TATA'. This improves accuracy over using only DMs or autoregressive models, as shown in \Cref{fig:front-page}.}
        \label{pic:max_logit_plot}
    \end{center}
    \vskip -0.2in
\end{figure}
Latent Diffusion Models (LDMs) may produce errors at the single nucleotide level during conversion from latent to input space.  We hypothesise that these inaccuracies are reflected in lower softmax coefficients output by the model. As illustrated in Figure \ref{pic:max_logit_plot}, these lower coefficients often cluster, forming noticeable `valleys' in regions of the sequence with low probability.

To address this issue, we introduce the `Absorb-Escape' algorithm, designed specifically to refine these problematic `valley' areas. The intuition behind Absorb-Escape is that while the Latent Diffusion Models effectively capture the overall structure of the sequences, they tend to make mistake at a single nucleotide level. 

This algorithm functions by detecting and amending these small-scale errors, thereby enhancing the overall accuracy of the generated sequences. For this refinement process, we utilize pre-trained autoregressive models, which are particularly effective in capturing local nuances and details of the sequences. This approach enables a more precise correction of the errors made by LDMs, resulting in sequences with improved accuracy both in local and global contexts.

Given a sequence \( \b s\) generated by a diffusion model $\mathcal{D}$ and a pre-trained autoregressive model $\mathcal{M}$. We use $L(\mathcal{D},\b s_i)$ and $L(\mathcal{M},\b s_i)$ to represent the maximum softmax coefficient returned by two models at position $i$. The Absorb-Escape algorithm operates as follows, with absorb condition \( \mathcal{A} \) and escape condition $\mathcal{E}$ defined in \Cref{table:ab-condition}.

\begin{enumerate}
    \item \textbf{Absorb:} Iterate through each token \( s_i \) $\in$ \(\b s \). Check if the absorb condition \( \mathcal{A}(s_i) \) is met. If \( \mathcal{A}(s_i) \) is true, proceed to the next step.
    \item \textbf{Autoregressive Generation:} Use a pretrained autoregressive model \( \mathcal{M} \) to refine the sequence starting from position \( i \) where \( \mathcal{A}(s_i) \) is satisfied.
    \item \textbf{Escape:} Continue the generation process until the escape condition \( \mathcal{E}(s_{j}, \text{params}) \) is satisfied. Replace the low probability region in the original sequence \( \b s \) with the newly generated sequence from \( \mathcal{M} \).
\end{enumerate}

When implementing the Absorb-Escape algorithm, various conditions and parameters can be selectively adjusted. Typically, we opt for the natural escape conditions (\Cref{table:ab-condition}) by default. This choice enables the autoregressive model to iteratively enhance local inaccuracies, continuing this refinement process until the diffusion model reaches a state of confidence regarding its output.
\begin{table}[!t]
\centering
\caption{Absorb and Escape Conditions. In these conditions, \( T_{\text{absorb}} \) represents the predefined absorb threshold. The escape threshold is indicated by \( T_{\text{escape}} \), while \( T_{\text{random}} \) stands for the random threshold. Lastly, \( \text{max\_sub\_length} \) means the maximum subsequence length.}
\resizebox{\linewidth}{!}{
\begin{tabular}{l|ll}
\hline
\textbf{Type} & \textbf{Condition Name} & \textbf{Description} \\
\hline
Absorb & Threshold Condition & \( \mathcal{A}(s_i) = \text{true if } L(\mathcal{D},\b s_i) < T_{\text{absorb}} \) \\
\hline
\multirow{4}{*}{Escape} & Natural  & \( \mathcal{E}(s_{j}) = \text{true if } L(\mathcal{D},\b s_j) > L(\mathcal{M},\b s_j) \) \\
\cline{2-3}
 & Threshold  & \( \mathcal{E}(s_{j}, T_{\text{escape}}) = \text{true if } L(\mathcal{M},\b s_j) < T_{\text{escape}} \) \\
\cline{2-3}
 & Random  & \( \mathcal{E}(s_{j}, T_{\text{random}}) = \text{true if } \text{rand.uniform}(0, 1) > T_{\text{random}} \) \\
\cline{2-3}
 & Max Length  & \( \mathcal{E}(s_{j}, \text{max\_sub\_length}) = \text{true if } j > \text{max\_sub\_length} \) \\
\hline
\end{tabular}
}
\label{table:ab-condition}
\end{table}


\section{Results and Evaluation}
\label{sec:results}
In \Cref{sec:ablation}, we conduct an ablation study to understand the impact of different components in DiscDiff.
In \Cref{sec:unconditional-results}, we compare DiscDiff with the baseline diffusion models on unconditional generations.
\Cref{sec:conditional-results} shows a comparison between DiscDiff and the autoregressive foundation model Hyena on conditional generations.
We show that the use of the Absorb-Escape algorithm can allow the user to trade-off between desired motifs in generated sequences.

\subsection{Ablation Study}
\label{sec:ablation}
\begin{table}[!t]
    \centering
    \caption{Ablation study: Comparison of 
    different versions of DiscDiff on EPD-GenDNA-large.}
    \resizebox{\linewidth}{!}{
    \begin{tabular}{
            l
            c
            c
            c
            c
        }
        \toprule
        Encode       & Acc$\uparrow$  & {S-FID $\downarrow$} & $\text{Cor}_{\text{TATA}}\uparrow$ & {$\Delta$Div$\downarrow$} \\
        \midrule
        CNN-VAE      & 99.4\%  & \textbf{45.2}                 & \textbf{0.858 }                             & \textbf{4.2\%}            \\
        CNN-Attn-VAE & 98.8\% & 53.6                 & 0.531                              & 9.8\%            \\
        1D-Swin-VAE  & \textbf{99.6\%} & 82.1                 & 0.651                              & 7.5\%            \\
        \bottomrule
    \end{tabular}
    \label{table:ablation-vae}
    }
\vspace{-10pt}
\end{table}

For the latent diffusion model, one important design choice is the architecture of the VAE.
We train different VAE architectures on EPD-GenDNA-large and compare their performance on unconditional generation.
During the first stage of training, three encoders: CNN-VAE, CNN-Attn-VAE, and 1D-Swin-VAE are trained to convergence.

We use the reconstruction accuracy of the VAEs as an evaluation metric in \Cref{table:ablation-vae}, which is defined as
the percentage of the reconstructed sequence that is identical to the input sequence from the VAE.
As shown by the first column of \Cref{table:ablation-vae},
1D-Swin-VAE achieves the highest reconstruction accuracy, while CNN-Attn-VAE achieves the lowest reconstruction accuracy.

However, when evaluated on their sequence generation capability after using these VAEs in DiscDiff, CNN-VAE achieves the best performance as measured by
S-FID and $\text{Cor}_{\text{TATA}}$. We then visualize the latent space of the VAEs by projecting it to 2D using UMAP~\citep{mcinnes2018umap}.
We calculate the number of dimensions needed to explain the $95\%$ of the variance. It is evident that 1D-Swin-VAE needs more dimensions to reach the $95\%$ of the variance compared to CNN-VAE.
Intuitively, this shows that CNN-VAE forms a simpler latent space than 1D-Swin-VAE. We hypothesize that this simpler latent space formulation significantly eases the subsequent training of the diffusion model. Detail and the visualizations are presented in \Cref{app:ablation-vae}.

\subsection{Unconditional Generation}
\label{sec:unconditional-results}
\begin{table*}[ht]
    \centering
    \caption{A comparison of diffusion models on unconditional generation evaluated on EPD-GenDNA-small and EPD-GenDNA-large. Metrics include S-FID, $\text{Cor}_{\text{TATA}}$, and $\Delta$Div. The best and second-best scores are highlighted in bold and underlined, respectively.}

    \resizebox{0.67\linewidth}{!}{
    \begin{tabular}{
            l
            c
            c
            c
            c
            c
            c
            c
        }
        \toprule
                                          & \multicolumn{3}{c}{EPD-GenDNA-small} & \multicolumn{3}{c}{EPD-GenDNA-large}                                                                                                                                               \\
        \cmidrule(lr){2-4} \cmidrule(lr){5-7}
        Model                             & {S-FID $\downarrow$}                 & $\text{Cor}_{\text{TATA}}\uparrow$   & {$\Delta$Div $\downarrow$}                  & {S-FID $\downarrow$}           & $\text{Cor}_{\text{TATA}}\uparrow$ & {$\Delta$Div $\downarrow$}                  \\
        \midrule
        Random (Reference)                & 119.0                                & -0.241                               & 29.3\%                             & 106.0                          & 0.030                              & 13.0\%                             \\
        Sample from Training Set          & 0.509                                & 1.0                                  & 0                             & 0.100                          & 0.999                              & 0                             \\
        \midrule
        VAE                               & 295.0                                & -0.167                               & \underline{0.40\%}                             & 250.0                          & 0.007                              & 10.6\%                             \\
        BitDiffusion                      & 405                                  & 0.058                                & 44.9\%                             & 100.0                          & 0.066                              & 2.00\%                             \\
        D3PM (small)                      & 97.4                                 & 0.0964                               & 28.0\% & 94.5                           & 0.363                              &  12.8\% \\
        D3PM (large)                      & 161.0                                & -0.208                               & \textbf{0.10\%}                             & 224.0                          & 0.307                              & \textbf{0.10\%}                             \\
        DDSM (Time Dilation)              & 504.0                                & 0.897                                & 40.6\%                             & 1113.0                         & 0.839                              & 13.0\%                             \\
        DiscDiff (Ours)                   & \underline{57.4}            & \underline{0.973}           & 4.40\%                             & \underline{45.2}      & \underline{0.858}         & 4.20\%                            \\
        \midrule
        Absorb-Escape (Ours)              & \textbf{3.21}                                 & \textbf{0.975}                         & 5.70\%                             & \textbf{4.38}                           & \textbf{0.892}                              &  \underline{1.90\%}                             \\
        \bottomrule
    \end{tabular}
    }
    \label{table:unconditional}
    \vspace{-10pt}
\end{table*}

First, we compare the performance of DiscDiff and other models on unconditional generation for mammalian DNA sequences and present the results in \Cref{table:unconditional}. The training set consists of 65,000 unique DNA sequences. We compare our model against the state-of-the-art discrete diffusion model: DDSM~\citep{avdeyev2023dirichlet}, D3PM~\citep{austin2021structured},  and BitDiffusoin~\citep{chen2022analog}. For the autoregressive models used in Absorb-Escape, we fine-tune the pre-trained Hyena~\citep{nguyen2023hyenadna} on EPD-GenDNA-large and EPD-GenDNA-small. The training budget is fixed to 72 GPU-Hours on NVIDIA A100 for all models. Training details are found in \Cref{app:baselines}.

\paragraph{A Comparison to other Diffusion Models} 
We generated 50,000 samples from each model.
Metrics measuring the generated sequences are detailed in  \Cref{table:unconditional}.
DiscDiff produces the most realistic DNA sequences compared to other diffusion baselines
on both short and long-sequence generations. This is indicated by both the smallest latent distance (S-FID) and largest motif frequency correlation ($\text{Cor}_{\text{TATA}}$) scores when compared to the ground truth. For diversity, we want the generated sequence to have similar diversity to the training data. This means that generated sequences have diversity that matches natural diversity. We measure it by $\Delta\text{Div}$, the diversity difference between natural and generated sequences. We see that D3PM (Large) has the smallest  $\Delta\text{Div}$ but performs worse than DiscDiff when it comes latent distance and motif frequency correlation.

\paragraph{The Absorb-Escape Algorithm}
The Absorb-Escape algorithm, which refines generated sequences of DiscDiff with Hyena, boosts the quality of generated sequences: it significantly reduces the S-FID and achieves the highest correlation score among all the models. We also compare our model against Hyena (an autoregressive model) in \Cref{app:extra-results}.

\subsection{Conditional Generation Results}
\label{sec:conditional-results}

\begin{figure}[!t]
    \begin{center}
        \centerline{\includegraphics[width=\columnwidth]{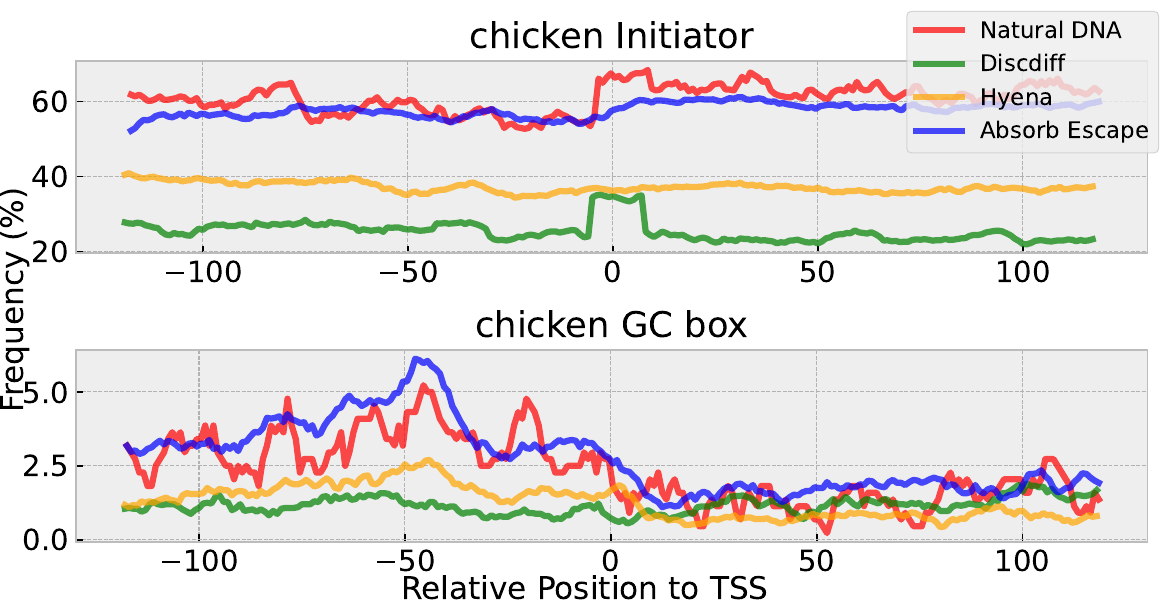}}
        \caption{Model Comparison of Chicken DNA Motif Distributions. This illustrates the Initiator and GC box frequencies across natural and generated DNA sequences near the TSS.}
        \label{fig:chicken}
    \end{center}
    \vskip -0.2in
\end{figure}

\begin{table*}[ht]
    \centering

    \caption{A performance comparison of models Hyena, DiscDiff, and Absorb-Escape on EPD-GenDNA-small for Baking Yeast and Fruit Fly. Models are evaluated on $\text{Cor}_{\text{TATA}}$ and $\text{Cor}_{\text{Initiator}}$. Best scores are highlighted and averages across 15 species are provided. The species-specific results are presented in \Cref{table:conditional-extra-results-1,table:conditional-extra-results-2,table:conditional-extra-results-3,table:conditional-extra-results-4,table:conditional-extra-results-5}.}
    \resizebox{0.8\linewidth}{!}{
    \begin{tabular}{
            l
            c
            c
            c 
            c
            c
            c
            c
            c
            c
        }
        \toprule
                      & \multicolumn{2}{c}{Baking Yeast}   & \multicolumn{1}{c}{...}                 & \multicolumn{2}{c}{Fruit Fly} & \multicolumn{3}{c}{\textbf{Average} (15 Species)}                                                                                                                                                                                                       \\ 
        \cmidrule(lr){2-3} \cmidrule(lr){4-4} \cmidrule(lr){5-6} \cmidrule(lr){7-9}
        Model         & $\text{Cor}_{\text{TATA}}\uparrow$ & $\text{Cor}_{\text{Initiator}}\uparrow$ &                               & $\text{Cor}_{\text{TATA}}\uparrow$                & $\text{Cor}_{\text{Initiator}}\uparrow$ & $\text{Cor}_{\text{TATA}}\uparrow$ & $\text{Cor}_{\text{Initiator}}\uparrow$ & $\frac{\text{Cor}_{\text{TATA}}+\text{Cor}_{\text{Initiator}}}{2}\uparrow$ \\ 
        \midrule
           Hyena   & \textbf{0.958}         & 0.723                                   & ...                           & \textbf{0.970}                        & 0.198                                   & \textbf{0.939}         & 0.488                                   & 0.714                                                                      \\ 
        DiscDiff         & 0.858                              & 0.824                                   & ...                           & 0.813                                             & \textbf{0.926}              & 0.894                              & \textbf{0.581}              & 0.738                                                                      \\ 
        Absorb-Escape & 0.864                              &\textbf{0.902}              & ...                           & 0.937                                             & 0.630                                   & 0.917                              & 0.568                                   & \textbf{0.743}                                                 \\ 
        \bottomrule
    \end{tabular}
    }
    \label{table:conditional}
    \vspace{-10pt}
\end{table*}

This section highlights the performance of DiscDiff in conditional DNA sequence generation. As shown in \Cref{fig:chicken}, we apply the Absorb-Escape algorithm and see that it improves DiscDiff's performance, particularly DiscDiffs's ability to generate DNA with desired motifs. Using a training set of 160,000 DNA sequences from 15 species, DiscDiff and Hyena were trained to generate 4,000 samples per species. We looked at the frequency and position of key genetic motifs in the generated samples (TATA-box, Initiator, GC-content, CCAAT-box). The full results are shown in \Cref{fig:motif-distributions1} to \Cref{fig:motif-distributions15} in \Cref{app:appendix_motif_distributions}.
When compared to Hyena, we saw that DiscDiff Escape algorithm was able to achieve better accuracy in trend replication of motif distributions. 
Employing the Absorb-Escape algorithm improves the quality of generated samples. The algorithm merges the strengths of both models and produces realistic samples in terms of the trend line and the frequency values as shown in \Cref{fig:chicken} -- Absorb-Escape (blue) closely tracks the Natural DNA frequency distribution (red).

DiscDiff and Hyena are better suited for modelling different types of motifs. \Cref{table:conditional} shows that DiscDiff outperforms Hyena in modelling Initiator motifs, while Hyena outperforms DiscDiff in modelling TATA-box, as
indicated by $\text{Cor}_{\text{TATA}}$ and $\text{Cor}_{\text{Initiator}}$.
Absorb-Escape combines the strength of both models, achieving the best performance in $\frac{\text{Cor}_{\text{TATA}}+\text{Cor}_{\text{Initiator}}}{2}$.
Species-specific results can be found in \Cref{table:conditional-extra-results-1} to \Cref{table:conditional-extra-results-5}.

\begin{figure}[ht]
    \begin{center}
        \centerline{\includegraphics[width=0.9\columnwidth]{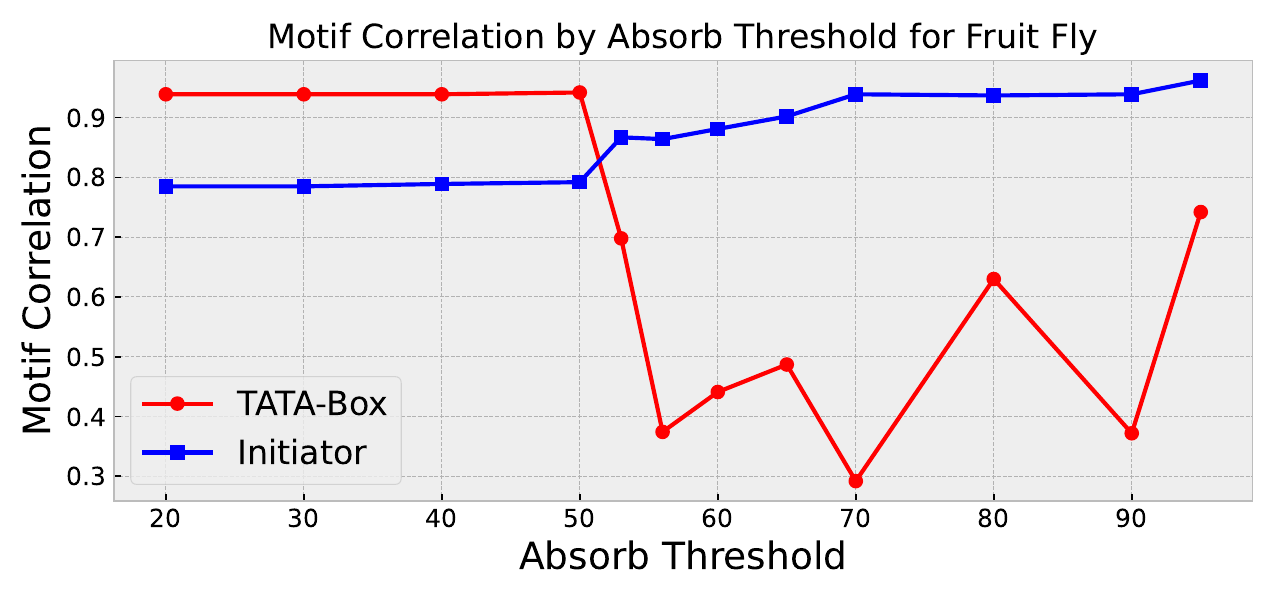}}
        \caption{Tuning genetic motif preferences with varied absorbance thresholds, \( T_{\text{absorb}} \). With the increment of \( T_{\text{absorb}} \), the  \( \text{Cor}_{\text{Initiator}} \) for sequences refined by the Absorb-Escape shows an increase, while \( \text{Cor}_{\text{TATA}} \) fluctuates. An optimal \( T_{\text{absorb}} \) can be selected to achieve high values for both \( \text{Cor}_{\text{TATA}} \) and \( \text{Cor}_{\text{Initiator}} \).}
        \label{fig:Absorb-Escape-trade-off}
    \end{center}
    \vskip -0.2in
\end{figure}

\paragraph{Balancing Motifs with Absorb-Escape}
The Absorb-Escape algorithm offers control over the properties of generated sequences. \Cref{fig:Absorb-Escape-trade-off} shows that varying the absorb threshold biases the generation of sequences towards a  preference for either the initiator or TATA-box motif distribution. This adaptability highlights the algorithm's utility in fine-tuning the balance between different genetic motifs in the generated sequences.

\section{Conclusion}
\label{sec:conclusion}
In this study, we have introduced a cross-species DNA dataset (EPD-GenDNA). This establishes a new benchmark for generative models in DNA sequence synthesis. Our conditional DNA model has demonstrated its proficiency in producing species-specific sequences using this dataset.

From a methodological perspective, we developed DiscDiff, a Latent Diffusion Model (LDM) specifically tailored for DNA generation. DiscDiff performs well on both conditional and unconditional DNA sequence generation tasks. Additionally, our Absorb-Escape algorithm effectively corrects mistakes made by diffusion models, significantly boosting the performance of DiscDiff on DNA generation. The versatility of our DiscDiff model extends beyond its current application, presenting the potential for testing in other types of discrete data and biological sequence generation tasks. 

The Absorb-Escape algorithm demonstrates potential in harnessing the advantages of both diffusion and autoregressive models. We hope that researchers will leverage this to enhance sequence generation performance in tasks involving discrete data.

\bibliography{reference}
\bibliographystyle{icml2024}

\newpage
\appendix
\onecolumn
\section{Dataset Preparation}
\label{appendix:dataset}

\begin{table}[htbp]
    \caption{The statistics of EPDnew dataset. }
    \label{table:epdnew}
    \centering
    \begin{tabular}{lll}
        \toprule
                                    & \# Unique Values & Examples                                                      \\
        \midrule
        genes                       & 130,014          & \text{CAACAAGGCACAGCAC...}                                    \\
        promoters                   & 159,125          & \text{GGCCCGGTTTTTTAAT...}                                    \\
        cell types                  & 2713             & [\text{Esophageal Epithelial Cells}, \text{Hippocampus}, ...] \\
        gene function description   & 130,014          & [\text{AHSP:alpha hemoglobin stabilizing protein}, ...]       \\
        species                     & 15               & [\text{Homo sapiens (Human).},\text{Zea mays (corn).}, ...]   \\
        expression levels           & 29,349,475       & [250,264,338, ...]                                            \\
        transcription starting site & 29,349,475       & [10,0,-1, ...]                                                \\
        \bottomrule
    \end{tabular}
\end{table}

\paragraph{Source Description}
The European Promoter Database (EPD) is a vital tool for biologists, with a primary focus on gathering and categorising promoters of eukaryotes. The EPD comprises promoter data sourced from numerous published articles. The EPDnew database, or High-Throughput EPD (HT-EPD), provides an even more comprehensive and rigorous dataset. The EPDnew has been enhanced by combining conventional EPD promoters with in-house analyses of promoter-specific high-throughput data, albeit for only selected organisms. This strategy enhances the precision and commendable coverage of EPDnew. It's worth noting that the evidence used in EPDnew primarily arises from Transcription Start Site (TSS) mapping derived from high-throughput experiments, notably CAGE and Oligocapping.

\paragraph{Data Selection}
Data was obtained from the EPDnew database without any exclusions, allowing for the incorporation of all available samples and associated promoters. Sequences between -1024 and +1023 base pairs from the transcription start site were chosen, ensuring a total length of 2048 base pairs - a convenient choice due to its alignment to powers of 2. This range covers crucial promoter elements like GC and TATA boxes, and also allows the algorithm to identify any possible hidden patterns and interconnections preceding and succeeding the transcription start site.

\paragraph{Time Frame}
The temporal aspect of our data is not limited to a specific time frame or year. The data was collected entirely as of August 2023.

\paragraph{Data collection and preprocessing}
Data collection started by manually downloading data files through FTP, followed by accessing fasta data via the website's graphical interface. Carefully extracted metadata from Python scripts was then integrated with promoter sequences to form a comprehensive table.

Given the species-specific fragmentation of the acquired data, a iterative method aided in bringing together diverse FASTA files into one consistent table, with important attributes such as 'kingdom' and 'species' integrated into the DataFrame at extraction. A crucial pre-processing step involved categorising DNA sequences as 'upstream' or 'downstream' based on their relative positioning. Normalisation and transformation techniques were implemented where necessary to ensure data uniformity and integrity were maintained.

Simultaneously, an additional dataset containing gene expression data underwent similar aggregation processes. Promoter expression sample data was synchronised with the main DataFrame. To optimize data storage and retrieval, the processed data was archived as a Pickle file.

Data integration and duplicate removal processes were also undertaken. Subsequently, the integrity and consistency of the data have been validated through multiple checks, including the verification of value counts for specific columns and a comparison of the data against EPDnew statistics.

\paragraph{Data Variables}

\begin{itemize}
    \item Promoter Sequences: -1024 to 1023 bp from tss
    \item PID: A unique identifier for each promoter in the EPDnew database.
    \item Species: one of 15 species in our dataset
    \item Average Expression Value: mean of expression values of every sample pertaining to a specified EPDnew promoter
    \item Sample Name:  Encompasses details like cell line and environmental conditions.
    \item Gene Description: description of gene function of the gene the promoter is regulating
\end{itemize}

\paragraph{Data Size and Structure}
As shown in Table \ref{table:epdnew}, the data set we compiled from EPDnew comprises of 159,125 distinct promoter sequences derived from 15 eukaryotic species, which are connected to 130,014 individual genes. Most of these genes have gene function definitions associated with them in the gathered data. The promoters have experimental expression levels and co-determined transcription start sites associated with them. These are frequently recorded in various cell types per promoter and might hold many experimental values for a single cell type. Our methodically structured dataset is in tabular form, facilitating smooth and comprehensive accessibility and manipulation for in-depth analysis.

\section{Justification of Two-Stage Training}
\label{app:justification}

The separation of VAE and the denoising model training can be justified
by the loss function of LSGM~\citep{vahdat2021score}.
LSGM jointly trains a VAE and the denoising model with a training loss consisting of a reconstruction
term and KL divergence between encoder distribution $p(z_0|x)$ and the prior distribution of the latent variable $p(z_0)$.
The latter term can be expressed in terms of the score function $\mathbf{S_\theta}(x(t),t)$. The optimisation of this loss is challenging,
requiring different weighting
mechanisms for the training objective and variance reduction techniques for stability.
The separation of training is equivalent to a special training schedule with the KL divergence term being fixed in the first stage, and then training the denoising model in the second stage. The performance gain of separating the training process has been shown in the image generation domain~\citep{rombach2022high,preechakul2022diffusion}.
In the section below, we provide the empirical evaluation of this method for DNA sequence generation.

\section{U-Net Configuration}
\label{app:unet-config}

U-Net is trained by sampling $t$ from uniform distribution $U[1,T]$ and $\b\varepsilon$ from $\mathcal{N}(\b 0,\b I)$.
We adopted the U-Net backbone from LDM~\citep{ho2204video,liu2023audioldm,ronneberger2015u} to implement  $\b\varepsilon_{\b\theta}(\b z_t,t)$.
Cross-attention~\citep{rombach2022high} is added to each layer to enable the integration of conditional information,
and self-attention~\citep{vaswani2017attention} is added to the middle layers to improve the expressive power of the network.

The UNet model used consists of four down and four up blocks. Each block consists of eight sequential ResNet blocks, followed by a single upsampling or downsampling block to change the number of channels. The third down block and the second up block are enriched with cross-attention layers, one applied after each ResNet block.

Each ResNet block transforms an input tensor through operational layers including normalisation, swish non-linearity, upsampling, two convolutional layers and time-embedding projection, culminating in an output tensor formed by summing the input tensor and the output of the second convolutional layer.

Attention layers, integral to the refinement of the feature representation, are incorporated in the third down and second up blocks, each consisting of eight 64-dimensional attention heads, executed by a scaled dot product attention mechanism without dropout and bias in the projection layers.

The UNet input and output contain 16 channels, each containing 16 x 16 arrays. The down blocks have channel dimensions of [256,256,512,512] and the up blocks mirror the encoder blocks in reverse order. In the forward process, we use N = 1000 steps. For EPD-GenDNA small, we reduce the embedding size to 8 x 8 x 8, and the other components of the U-Net are reduced proportionally.
\section{Ablation VAE}
\label{app:ablation-vae}
\subsection{VAE variants}
\paragraph{CNN-VAE}

\label{appendix:vae}

\begin{figure}[h!]
\centering
\includegraphics[width = 0.7\textwidth]{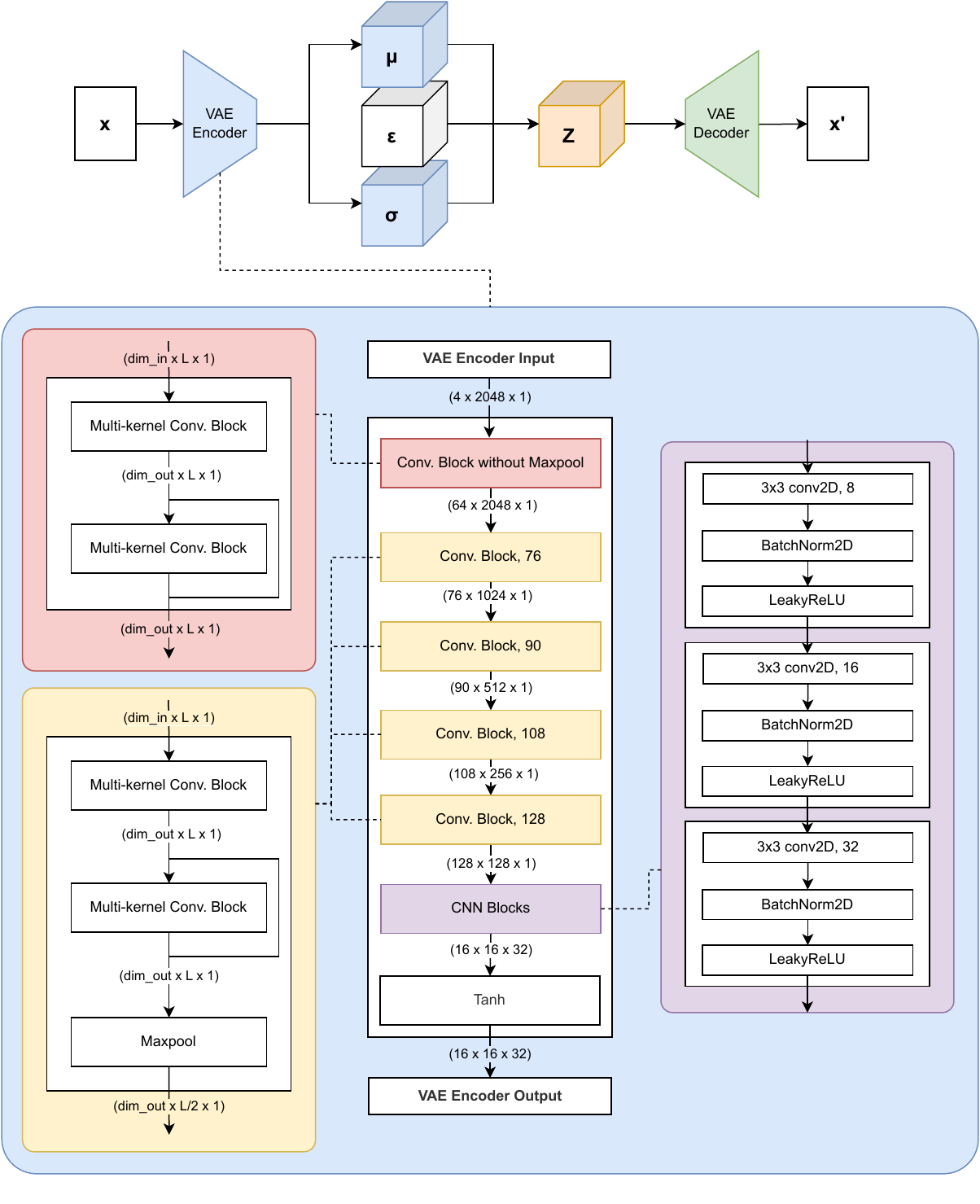}
\caption{\textbf{VAE Encoder Architecture}}
Two sections can be distinguished from the architecture: 1) the discrete data encoder (in red and yellow) and 2) the CNN Blocks (in purple)
\label{figure:appendix_vae2}
\end{figure}

The convolutional VAE follows the encoder-decoder architecture. The encoder learns how to encode the discrete DNA sequence data into a continuous latent representation. Simultaneously, the decoder learns how to decode the latent representations into DNA sequence data in continuous space before it is quantised with ArgMax. A detailed illustration of the encoder architecture is in figure \ref{figure:appendix_vae2}.

The decoder employs an architecture in symmetry with the encoder. In the decoder the same network layers are performed in reverse order and inverse operations are used in place of forward operations. Such inverse operation pairs include transposed convolution and convolution operation, upsampling and maxpooling operation.

\newpage
The first section of the encoder encodes the one-hot encoded input into a 2-dimensional image-like representation. From the initial 4 channels due to one-hot encoding, the channel dimension is first increased to 64, then it increases further with steps of exponentially increasing number of kernels up to 128. This is implemented using Multi-kernel Convolution Blocks with perturbable kernel parameters combined with residual connections. By including a max-pooling layer in each layer, its length shrinks by powers of 2 to 128. This effectively expands our 4x2048x1 data into 128x128x1, an image-like surface when viewed along the channel and length. This stage of the encoder aims to capture high-level positional features along the length dimension via pooling, for each channel dimension of the DNA sequence by transforming from 1-dimensional features into a 2-dimensional surface.

The second section of the encoder uses a CNN architecture for learning latent representations. There are three layers of the CNN block, each performing 3x3 kernel on the channel-length surface followed by BatchNorm2D and LeakyReLU activation. The CNN block in the encoder learns to produce different feature maps of size 16x16 from the 128x128 surface, which are concatenated into a 3-dimensional latent representation of shape 16x16x32, effectively increasing the width dimension. The latent representation can be partitioned into two blocks of size 16x16x16, representing the mean and log-variance of the learned features. Assuming priors of diagonal Gaussian distributions, we can take samples from the standard Gaussian distribution and use the reparameterisation trick to optimise for the variational learning. 

\begin{figure}[h!]
\centering
\includegraphics[width = 0.45\textwidth]{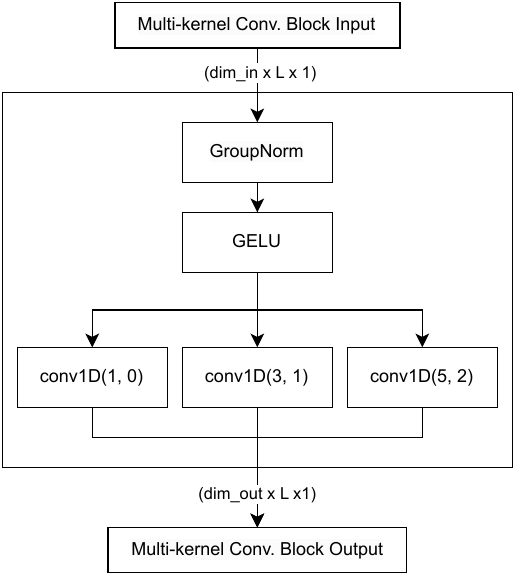}
\caption{\textbf{An Example of Multi-kernel Convolution Block}}
\label{figure:appendix_mult_kernel_conv}
\end{figure}

The Multi-kernel Convolution Block is the elementary building block in the first section of the VAE network. The block is structured to first perform normalisation and activation first, then conv1D operations are performed on a list of kernel sizes, where the results are then concatenated. An example Multi-kernel Convolution Block with the list of kernel sizes [1, 3, 5] is illustrated in figure \ref{figure:appendix_mult_kernel_conv}.

\paragraph{CNN-Attn-VAE}
For CNN-Attn-VAE, we replace the first stage encoder of the VAE encoder with CNNs and self-attention layers as used in Enformer~\citep{avsec2021effective} for genomic assay prediction.

\paragraph{1D-Swin-VAE} 
For 1D-Swin-VAE, we replace the first stage encoder of the VAE encoder with 1-d Swin~\citep{li2023genomic}, a type of sparse transformer for genomic sequence modelling.

\subsection{Embedding Space Analysis}

\begin{figure}[h!]
\centering
\includegraphics[width = 0.45\textwidth]{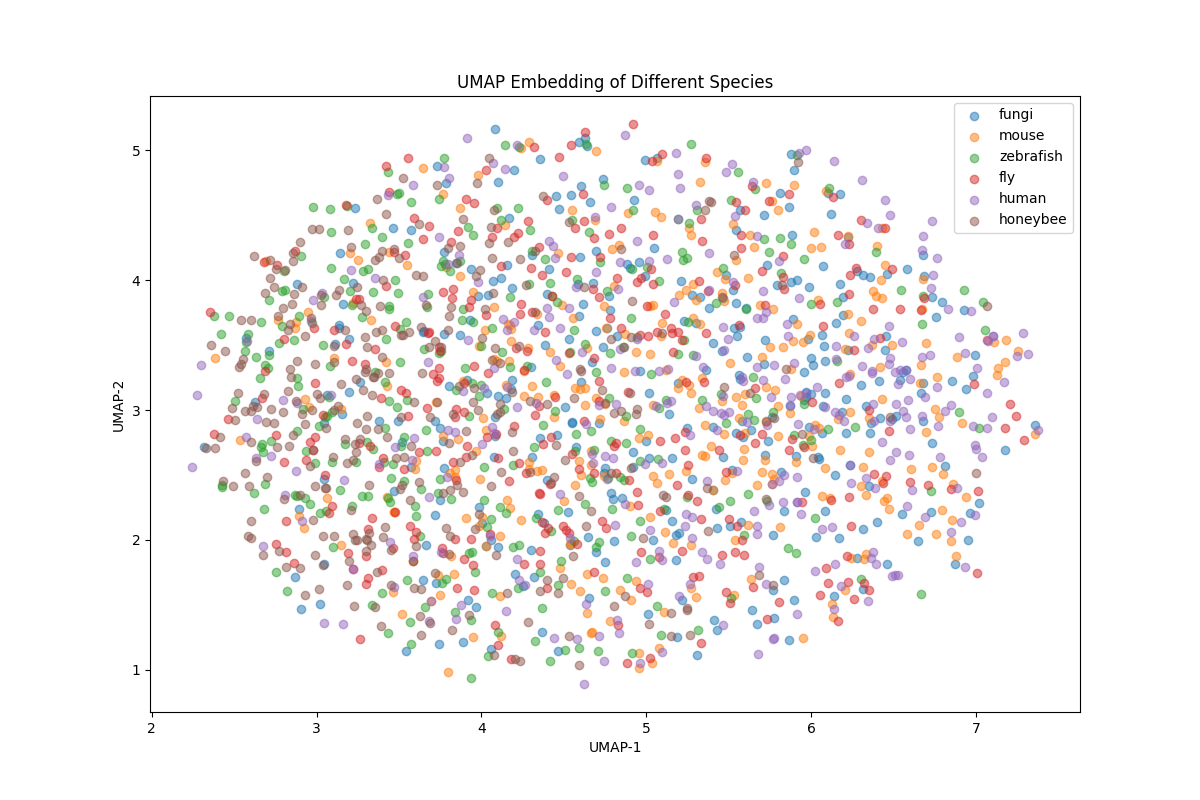}
\includegraphics[width = 0.45\textwidth]{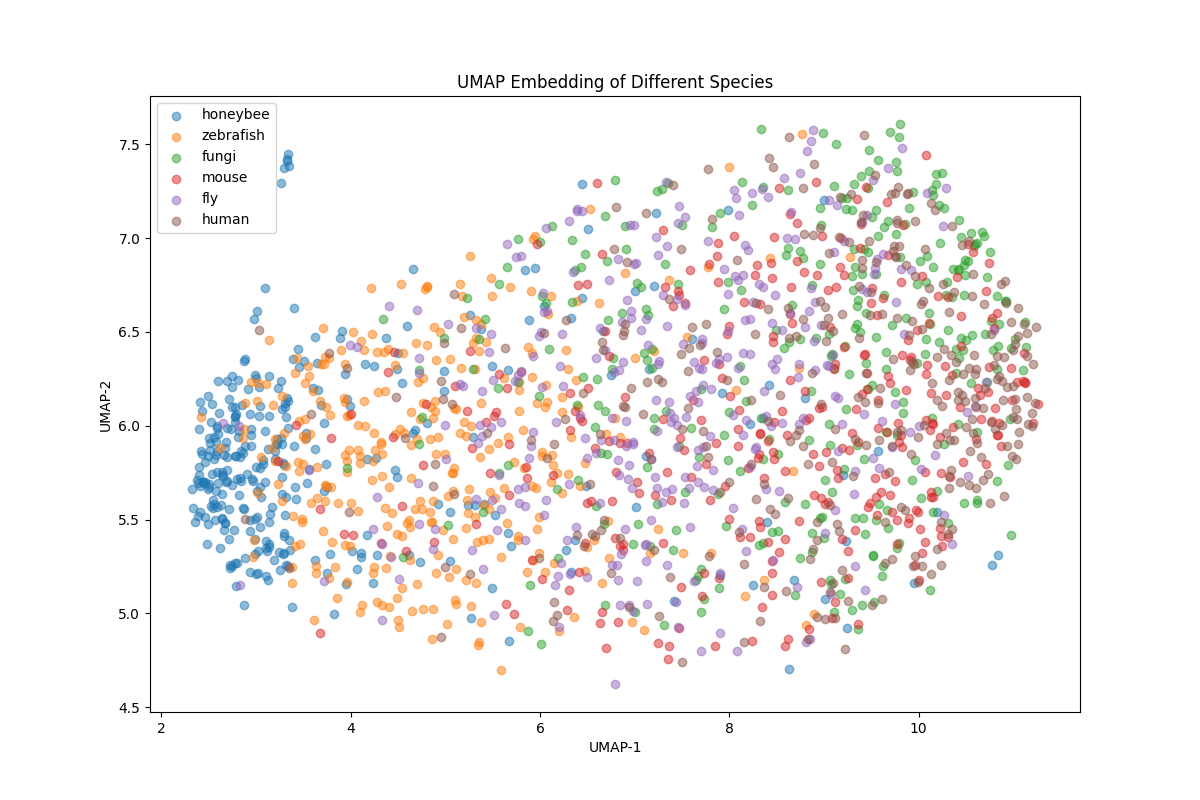}
\includegraphics[width = 0.45\textwidth]{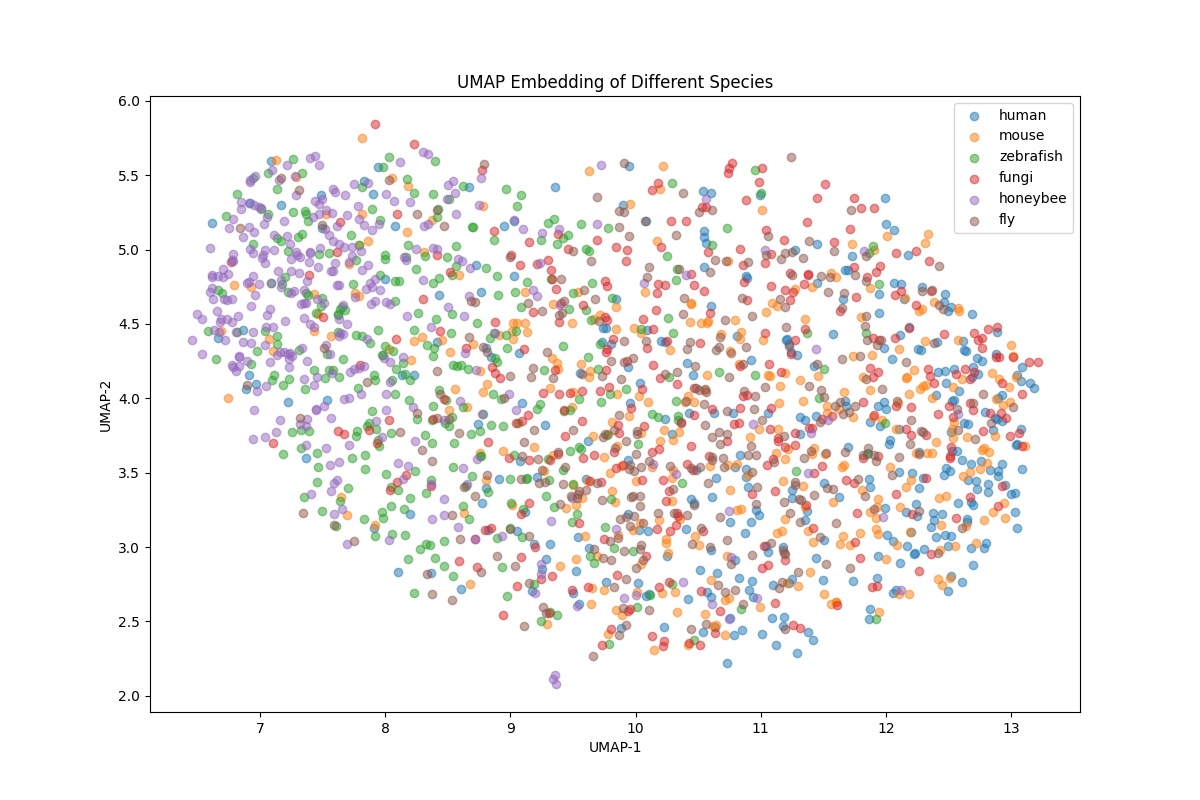}
\label{figure:appendix_figure_embedding}
\caption{The top one is the latent space of CNN-VAE, while the bottom left is for CNN-Attn-VAE and the bottom right is 1D-Swin-VAE, respectively.}
\end{figure}

Here, we explore the embedding space of three encoders. \Cref{figure:appendix_figure_embedding}  shows the latent distance embedding of three encoders with UMAP~\citep{mcinnes2018umap}. Note that while the species information is not given during the training of VAE, it can be seen that while the data points in the embedding space of 1D-Swin-VAE and CNN-Attn-VAE automatically cluster into species, the embedding space from CNN-VAE doesn't cluster. While this behaviour could be useful for unsupervised learning, it has the potential to increase the complexity of the learnt embedding space. Hereby, it may make the learning process in the second stage of LDM more challenging, resulting in lower-quality examples. The complexity of the embedding space can also be measured by the number of components required to recover 95\% of the variance in the original dataset with PCA. For two models: 1D-Swin-VAE and CNN-VAE which archive relatively the same amount of reconstruction accuracy. CNN-VAE requires 1312 components to recover the 95\% variance, while 1D-swin-VAE requires 1342 components. 
\section{DiscDiff and Baseline Models}
\label{app:baselines}
\paragraph{DiscDiff Training}

The model was trained utilizing the NVIDIA RTX A6000 graphics card. During the first stage, a Variational Autoencoder (VAE) was trained on a dataset comprising 160,000 sequences spanning 15 different species. This training was executed over 24 GPU hours with the primary goal of accurately reconstructing the input sequences. The chosen learning rate was \(0.0001\), paired with the Adam Optimizer. A batch size of \(128\) was set, and the KL divergence weight was determined at \(10^{-4}\). The resultant hidden representation, \(z\), belongs to \(\mathbb{R}^{16 \times 16\times 16}\), which is precisely half the dimensionality of the original input \(x \in \mathbb{R}^{4 \times 2048}\).

In the second training phase, a denoising UNet and other baselines were trained on NVIDIA A100 40GB  with a fixed budget of 72 GPU-Hours. Given the GPU RAM capacity of \(40\) GB, we optimized the model by maximizing the number of residual blocks and self-attention layers. The training also incorporated a cosine learning scheduler, introducing a warm-up period for the initial \(500\) epochs. The learning rate was set at \(0.00005\), and a batch size of \(256\) was chosen for the training. DDPM~\citep{ho2020denoising} is used for sampling. 

For a focused evaluation of the model's performance on unconditional generation, particularly in mammalian DNA sequence generation, we opted for a subset of the entire dataset. This subset includes sequences from \textit{H. Sapiens} (human), \textit{Rattus Norvegicus} (rat), \textit{Macaca mulatta}, and \textit{Mus musculus} (mouse), which collectively represent 50\% of the total dataset. Training on this subset not only streamlines the evaluation process but also allows for a more precise assessment of the generative algorithm's efficacy and accuracy in producing mammalian sequences

\paragraph{Baselines} The details about the \textbf{architecture and implementation} of the baseline models are as below:
\begin{itemize}
\item BitDiffusion~\citep{chen2022analog}: We enhance the current BitDiffusion implementation for DNA synthesis, originally from the DNA-Diffusion project\footnote{\url{https://github.com/pinellolab/DNA-Diffusion}}, by expanding the models to encompass 380 million parameters. This network is composed of Convolutional Neural Networks (CNNs), interspersed with layers of cross-attention and self-attention.

\item DDSM~\citep{avdeyev2023dirichlet}: We scale up the original implementation of the denoising network used for promoter design in DDSM\footnote{\url{https://github.com/jzhoulab/ddsm}} to what is the corresponding size of the network given 470 million parameters. It is a convolution-based architecture with dilated convolution layers.
\item D3PM~\citep{austin2021structured}: We take the implementation of D3PM for biological sequence generation from EvoDiff~\citep{alamdari2023protein}\footnote{\url{https://github.com/microsoft/evodiff}}, adopting the algorithm for DNA generation. We use the original implementation of the denoising network, which has two versions: with sizes of 38M and 640M. We hereby have D3PM (small) and D3PM (big), respectively. 
\item Hyena~\citep{nguyen2023hyenadna}: We modify the RegLM~\citep{lal2023reglm}\footnote{\url{https://github.com/Genentech/regLM}}, a existing work uses hyena for DNA generation. Four pretrained Hyena models of different sizes (hyenadna-large-1m-seqlen, hyenadna-medium-160k0seqlen, heynadna-small-32k-seqlen, and hyenaana-tiny-16k-seqlen-d128) are downloaded from HuggingFace\footnote{\url{https://huggingface.co/LongSafari}} and used for full-size fine-tuning, we apply the fine-tuned models for generations on EPD-GenDNA.
    
\end{itemize}
\newpage
\section{Additional Results}
\label{app:extra-results}
We provide additional results for conditional generations on EPD-GenDNA-small dataset below. \Cref{table:unconditional-extra} provides the unconditional generation results for Hyena models together with other models. 
\begin{table*}[ht]
    \centering
    \begin{tabular}{
            l
            S[table-format=2.1]
            S[table-format=1.3]
            S[table-format=3.1]
            S[table-format=1.3]
            S[table-format=3.1]
            S[table-format=1.3]
        }
        \toprule
                      & \multicolumn{2}{c}{Baking Yeast} & \multicolumn{2}{c}{Chicken} & \multicolumn{2}{c}{Corn}                                                                        \\
        \cmidrule(lr){2-3} \cmidrule(lr){4-5} \cmidrule(lr){6-7}
        Model         & { TATA $\uparrow$}               & {Initiator $\uparrow$}      & { TATA $\uparrow$}       & {Initiator $\uparrow$} & { TATA $\uparrow$} & {Initiator $\uparrow$} \\
        \midrule
        Hyena         & 0.958                            & 0.723                       & 0.956                    & 0.553                  & 0.995              & 0.875                  \\
        DiscDiff      & 0.858                            & 0.824                       & 0.960                    & 0.683                  & 0.983              & 0.796                  \\
        Absorb-Escape & 0.864                            & 0.902                       & 0.910                    & 0.625                  & 0.991              & 0.824                  \\
        \bottomrule
    \end{tabular}
    \caption{Conditional Generations on EPD-GenDNA-small: for human, honee, and baking yeast.}
    \label{table:conditional-extra-results-1}
\end{table*}

\begin{table*}[ht]
    \centering
    \begin{tabular}{
            l
            S[table-format=2.1]
            S[table-format=1.3]
            S[table-format=3.1]
            S[table-format=1.3]
            S[table-format=3.1]
            S[table-format=1.3]
        }
        \toprule
                      & \multicolumn{2}{c}{Dog} & \multicolumn{2}{c}{Elegans} & \multicolumn{2}{c}{Fission Yeast}                                                                        \\
        \cmidrule(lr){2-3} \cmidrule(lr){4-5} \cmidrule(lr){6-7}
        Model         & { TATA $\uparrow$}      & {Initiator $\uparrow$}      & { TATA $\uparrow$}                & {Initiator $\uparrow$} & { TATA $\uparrow$} & {Initiator $\uparrow$} \\
        \midrule
        Hyena         & 0.890                   & 0.130                       & 0.930                             & 0.592                  & 0.929              & 0.320                  \\
        DiscDiff      & 0.829                   & 0.167                       & 0.677                             & 0.947                  & 0.860              & 0.139                  \\
        Absorb-Escape & 0.754                   & 0.527                       & 0.857                             & 0.428                  & 0.947              & 0.412                  \\
        \bottomrule
    \end{tabular}
    \caption{Conditional Generations on EPD-GenDNA-small: for human, honey bee, and baking yeast.}
    \label{table:conditional-extra-results-2}
\end{table*}

\begin{table*}[ht]
    \centering
    \begin{tabular}{
            l
            S[table-format=2.1]
            S[table-format=1.3]
            S[table-format=3.1]
            S[table-format=1.3]
            S[table-format=3.1]
            S[table-format=1.3]
        }
        \toprule
                      & \multicolumn{2}{c}{Fruit Fly} & \multicolumn{2}{c}{Honey Bee} & \multicolumn{2}{c}{Human}                                                                        \\
        \cmidrule(lr){2-3} \cmidrule(lr){4-5} \cmidrule(lr){6-7}
        Model         & { TATA $\uparrow$}            & {Initiator $\uparrow$}        & { TATA $\uparrow$}        & {Initiator $\uparrow$} & { TATA $\uparrow$} & {Initiator $\uparrow$} \\
        \midrule
        Hyena         & 0.970                         & 0.198                         & 0.933                     & 0.779                  & 0.922              & 0.263                  \\
        DiscDiff      & 0.813                         & 0.926                         & 0.897                     & 0.924                  & 0.908              & 0.579                  \\
        Absorb-Escape & 0.937                         & 0.630                         & 0.905                     & 0.771                  & 0.900              & 0.411                  \\
        \bottomrule
    \end{tabular}
    \caption{Conditional Generations on EPD-GenDNA-small: for human, honey bee, and baking yeast.}
    \label{table:conditional-extra-results-3}
\end{table*}

\begin{table*}[ht]
    \centering
    \begin{tabular}{
            l
            S[table-format=2.1]
            S[table-format=1.3]
            S[table-format=3.1]
            S[table-format=1.3]
            S[table-format=3.1]
            S[table-format=1.3]
        }
        \toprule
                      & \multicolumn{2}{c}{Macaque} & \multicolumn{2}{c}{Mouse} & \multicolumn{2}{c}{Plasmodium}                                                                        \\
        \cmidrule(lr){2-3} \cmidrule(lr){4-5} \cmidrule(lr){6-7}
        Model         & { TATA $\uparrow$}          & {Initiator $\uparrow$}    & { TATA $\uparrow$}             & {Initiator $\uparrow$} & { TATA $\uparrow$} & {Initiator $\uparrow$} \\
        \midrule
        Hyena         & 0.935                       & 0.641                     & 0.959                          & 0.463                  & 0.792              & -0.284                 \\
        DiscDiff      & 0.939                       & 0.700                     & 0.941                          & 0.628                  & 0.888              & -0.686                 \\
        Absorb-Escape & 0.949                       & 0.745                     & 0.930                          & 0.471                  & 0.853              & -0.419                 \\
        \bottomrule
    \end{tabular}
    \caption{Conditional Generations on EPD-GenDNA-small: for human, honey bee, and baking yeast.}
    \label{table:conditional-extra-results-4}
\end{table*}

\begin{table*}[ht]
    \centering
    \begin{tabular}{
            l
            S[table-format=2.1]
            S[table-format=1.3]
            S[table-format=3.1]
            S[table-format=1.3]
            S[table-format=3.1]
            S[table-format=1.3]
        }
        \toprule
                      & \multicolumn{2}{c}{Rat} & \multicolumn{2}{c}{Thale Cress} & \multicolumn{2}{c}{Zebrafish}                                                                        \\
        \cmidrule(lr){2-3} \cmidrule(lr){4-5} \cmidrule(lr){6-7}
        Model         & { TATA $\uparrow$}      & {Initiator $\uparrow$}          & { TATA $\uparrow$}            & {Initiator $\uparrow$} & { TATA $\uparrow$} & {Initiator $\uparrow$} \\
        \midrule
        Hyena         & 0.972                   & 0.492                           & 0.979                         & 0.918                  & 0.969              & 0.664                  \\
        DiscDiff      & 0.939                   & 0.618                           & 0.970                         & 0.697                  & 0.953              & 0.773                  \\
        Absorb-Escape & 0.975                   & 0.671                           & 0.994                         & 0.893                  & 0.985              & 0.594                  \\
        \bottomrule
    \end{tabular}
    \caption{Conditional Generations on EPD-GenDNA-small: for human, honey bee, and baking yeast.}
    \label{table:conditional-extra-results-5}
\end{table*}

\begin{table}[ht]
    \centering
    \begin{tabular}{
            l
            S
            S
            S
            S
        }
        \toprule
        Encode       & Acc  & {S-FID $\downarrow$} & $\text{Cor}_{\text{TATA}}\uparrow$ & {$\Delta$Div$\downarrow$} \\
        \midrule
        CNN-VAE      & 99.4\%  & 45.2                 & 0.858                              & 4.2\%            \\
        CNN-Attn-VAE & 98.8\% & 53.6                 & 0.531                              & 9.8\%            \\
        1D-Swin-VAE  & 99.6\% & 82.1                 & 0.651                              & 7.5\%            \\
        \bottomrule
    \end{tabular}
    \caption{Unconditional Generations: Comparison of different version of DiscDiff on EPD-GenDNA-large.}
    \label{table:ablation-vae-appendix}
\end{table}

\begin{table*}[ht]
    \centering
    \begin{tabular}{
            l
            S
            S
            S
            S
            S
            S
            S
        }
        \toprule
                                          & \multicolumn{3}{c}{EPD-GenDNA-small} & \multicolumn{3}{c}{EPD-GenDNA-large}                                                                                                                                               \\
        \cmidrule(lr){2-4} \cmidrule(lr){5-7}
        Model                             & {S-FID $\downarrow$}                 & $\text{Cor}_{\text{TATA}}\uparrow$   & {Div $\uparrow$}                  & {S-FID $\downarrow$}           & $\text{Cor}_{\text{TATA}}\uparrow$ & {Div $\uparrow$}                  \\
        \midrule
        Random (Reference)                & 119.0                                & -0.241                               & 0.828                             & 106.0                          & 0.030                              & 0.261                             \\
        Sample from Training Set          & 0.509                                & 1.0                                  & 0.535                             & 0.100                          & 0.999                              & 0.131                             \\
        \midrule
        VAE                               & 295.0                                & -0.167                               & 0.531                             & 250.0                          & 0.007                              & 0.025                             \\
        BitDiffusion                      & 405                                  & 0.058                                & 0.086                             & 100.0                          & 0.066                              & 0.111                             \\
        D3PM (small)                      & 97.4                                 & 0.0964                               & \cellcolor{purple!25}\fbox{0.815} & 94.5                           & 0.363                              & \cellcolor{purple!25}\fbox{0.259} \\
        D3PM (large)                      & 161.0                                & -0.208                               & 0.534                             & 224.0                          & 0.307                              & 0.132                             \\
        DDSM (Time Dilation)              & 504.0                                & 0.897                                & 0.129                             & 1113.0                         & 0.839                              & 0.001                             \\
        DiscDiff (Ours)                   & \cellcolor{purple!25}57.4            & \cellcolor{purple!25}0.973           & 0.491                             & \cellcolor{purple!25}45.2      & \cellcolor{purple!25}0.858         & 0.173                             \\
        \midrule
        \color{FadedBlack} Hyena (Tiny)   & \color{FadedBlack} 2.97              & \color{FadedBlack} 0.972             & \color{FadedBlack} 0.535          & \color{FadedBlack} 3.34        & \color{FadedBlack} \fbox{0.940}    & \color{FadedBlack} 0.141          \\
        \color{FadedBlack} Hyena (Small)  & \color{FadedBlack} 3.20              & \color{FadedBlack} 0.913             & \color{FadedBlack} 0.530          & \color{FadedBlack} \fbox{1.16} & \color{FadedBlack} 0.924           & \color{FadedBlack} 0.136          \\
        \color{FadedBlack} Hyena (Middle) & \color{FadedBlack} 2.13              & \color{FadedBlack} 0.769             & \color{FadedBlack} 0.538          & \color{FadedBlack} 2.54        & \color{FadedBlack} 0.925           & \color{FadedBlack} 0.141          \\
        \color{FadedBlack} Hyena (Large)  & \color{FadedBlack} \fbox{1.76}       & \color{FadedBlack} 0.952             & \color{FadedBlack} 0.558          & \color{FadedBlack} 1.51        & \color{FadedBlack} 0.887           & \color{FadedBlack} 0.138          \\

        \midrule
        Absorb-Escape (Ours)              & 3.21                                 & \fbox{0.975}                         & 0.478                             & 4.38                           & 0.892                              & 0.150                             \\
        \bottomrule
    \end{tabular}
    \caption{Unconditional Generations: Comparison of models on EPD-GenDNA-small and EPD-GenDNA-large. }
    \label{table:unconditional-extra}
\end{table*}

\newpage
\section{Motif Distributions for 15 species}
\label{app:appendix_motif_distributions}
We plot TATA-box, GC content, Initiator, and CCAAT-box for 15 speces as below.

\begin{figure}[h]
    \centering
    \includegraphics[width=0.8\textwidth]{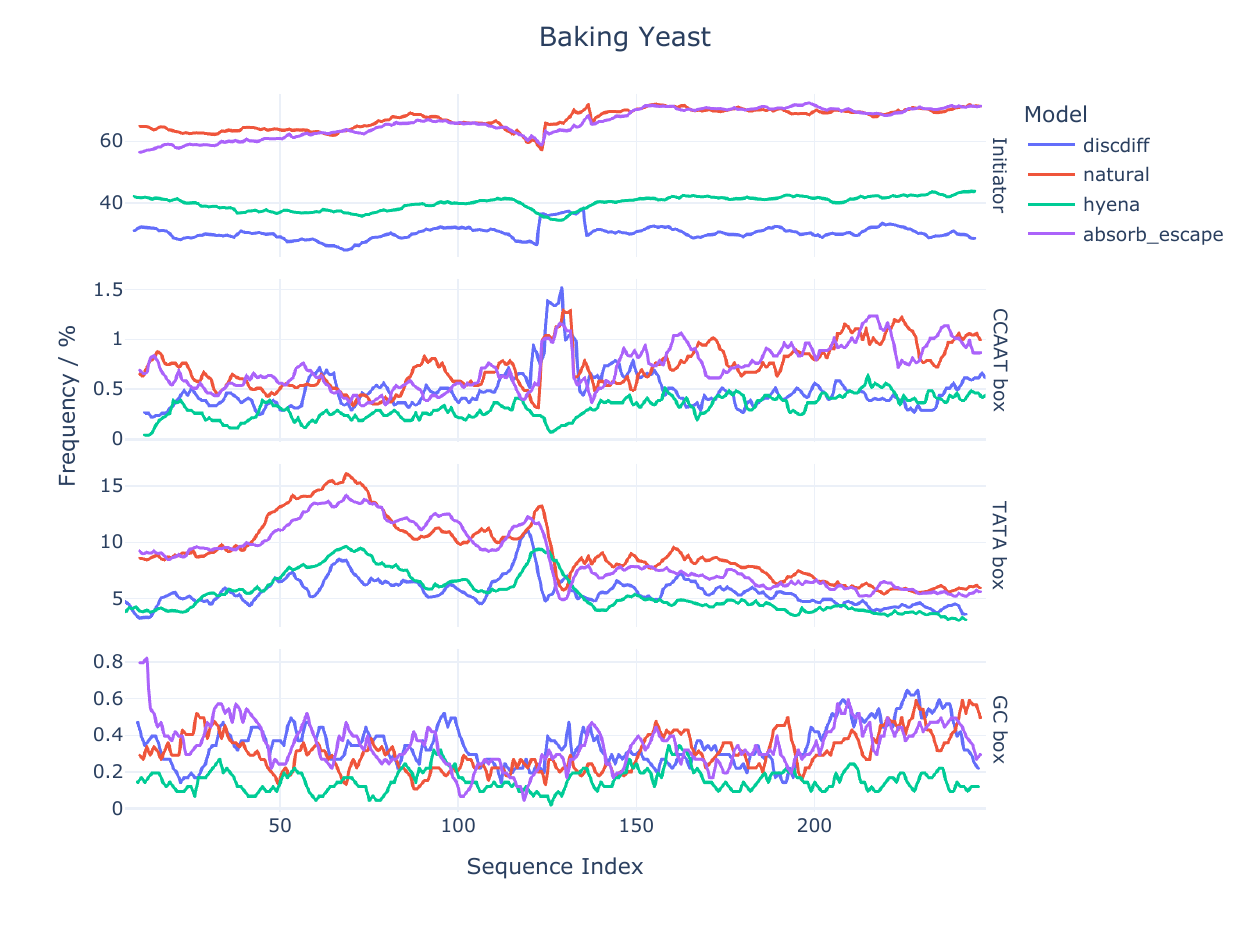}
    \caption{Baking Yeast}
    \label{fig:motif-distributions1}
\end{figure}

\begin{figure}[h]
    \centering
    \includegraphics[width=0.8\textwidth]{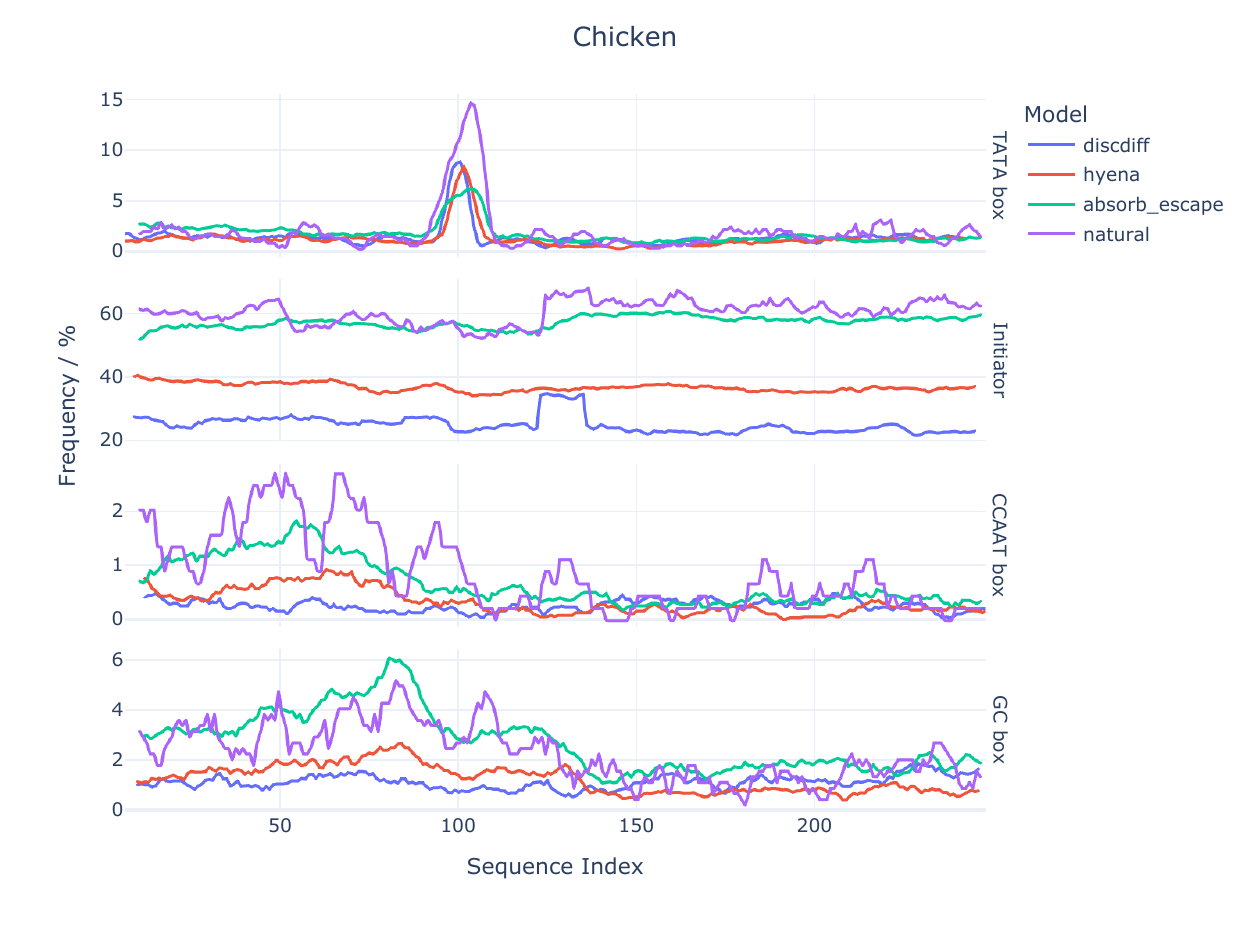}
    \caption{Chicken}
    \label{fig:motif-distributions2}
\end{figure}

\begin{figure}[h]
    \centering
    \includegraphics[width=0.8\textwidth]{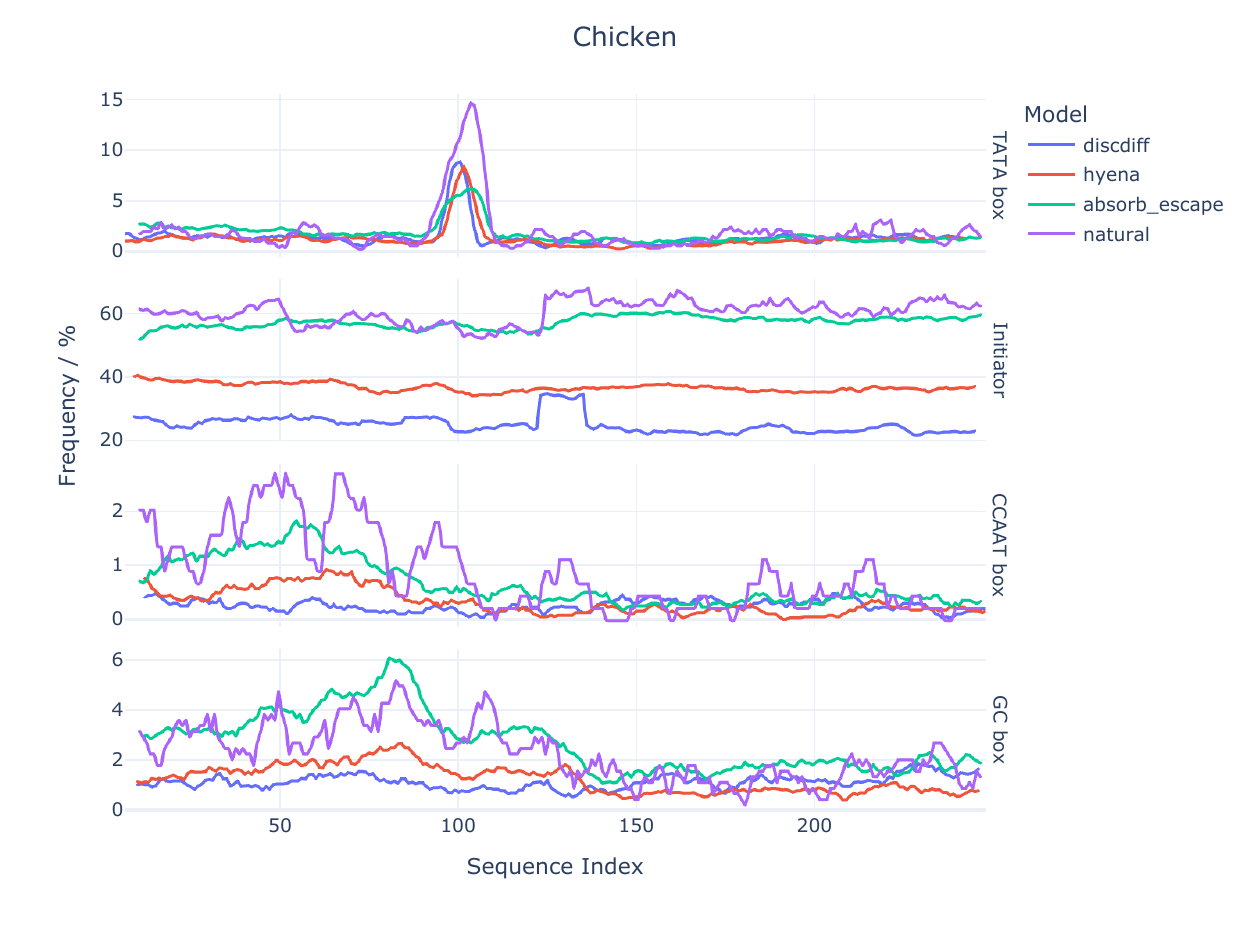}
    \caption{Chicken}
    \label{fig:motif-distributions3}
\end{figure}

\begin{figure}[h]
    \centering
    \includegraphics[width=0.8\textwidth]{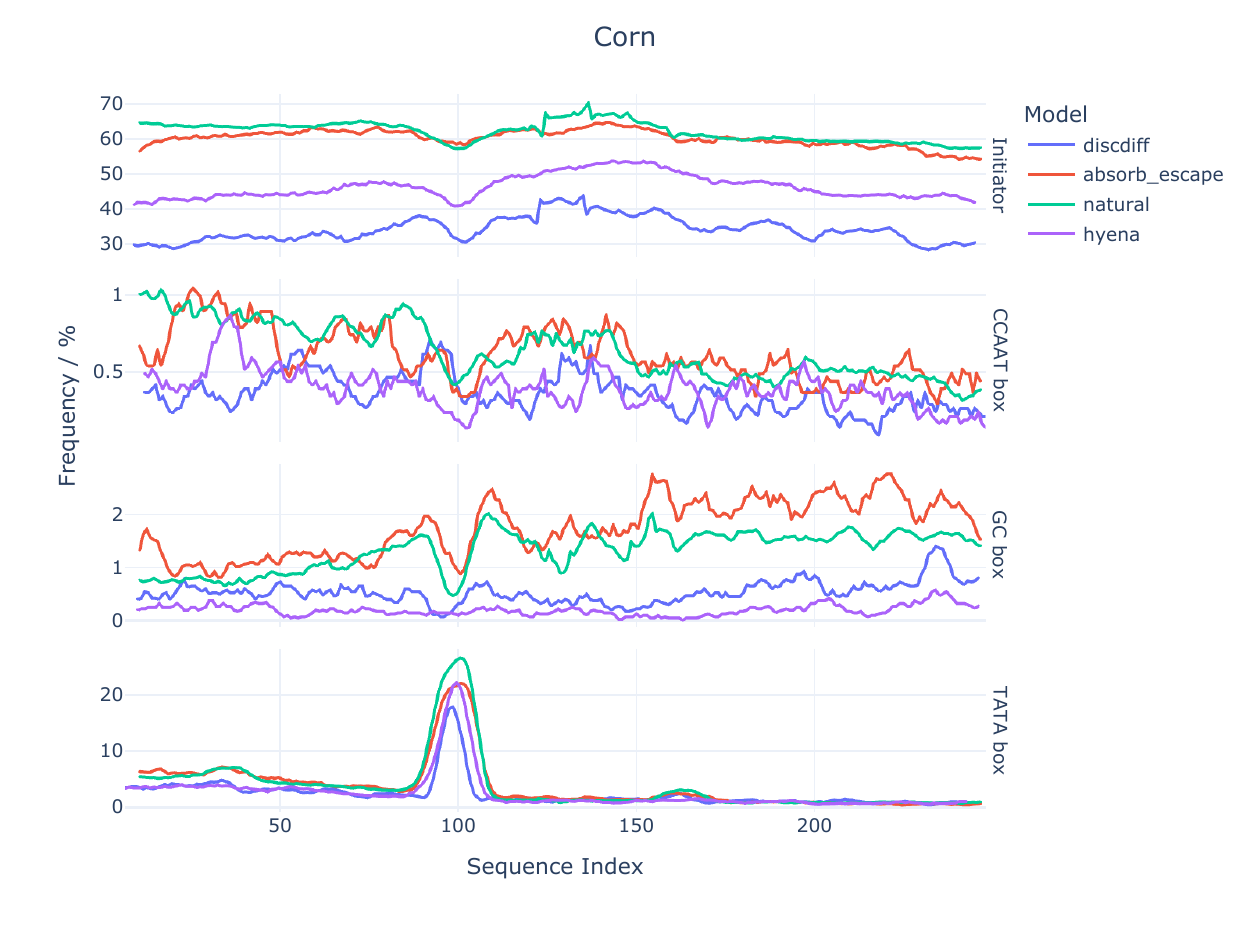}
    \caption{Corn}
    \label{fig:motif-distributions4}
\end{figure}

\begin{figure}[h]
    \centering
    \includegraphics[width=0.8\textwidth]{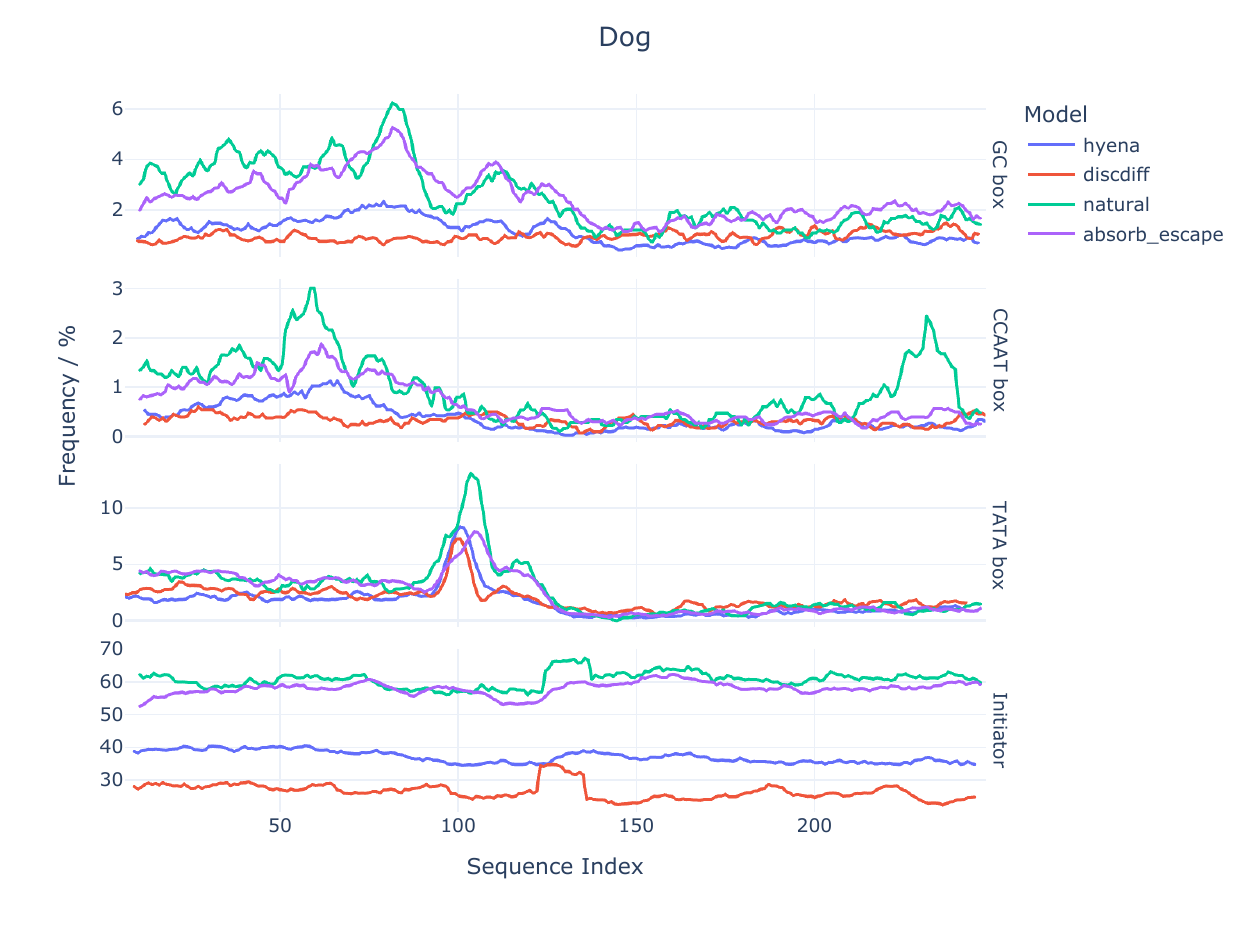}
    \caption{dog}
    \label{fig:motif-distributions5}
\end{figure}

\begin{figure}[h]
    \centering
    \includegraphics[width=0.8\textwidth]{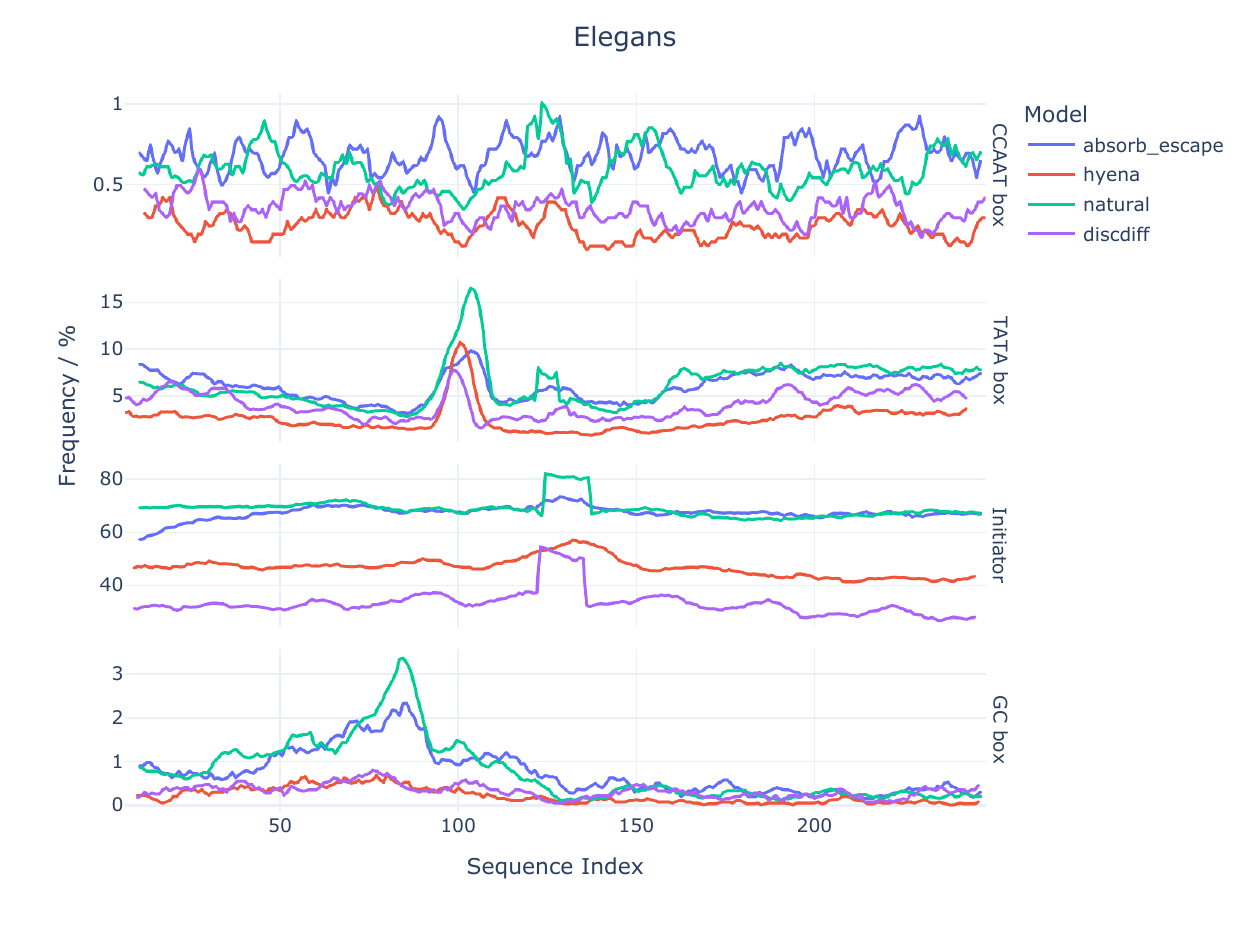}
    \caption{elegans}
    \label{fig:motif-distributions6}
\end{figure}

\begin{figure}[h]
    \centering
    \includegraphics[width=0.8\textwidth]{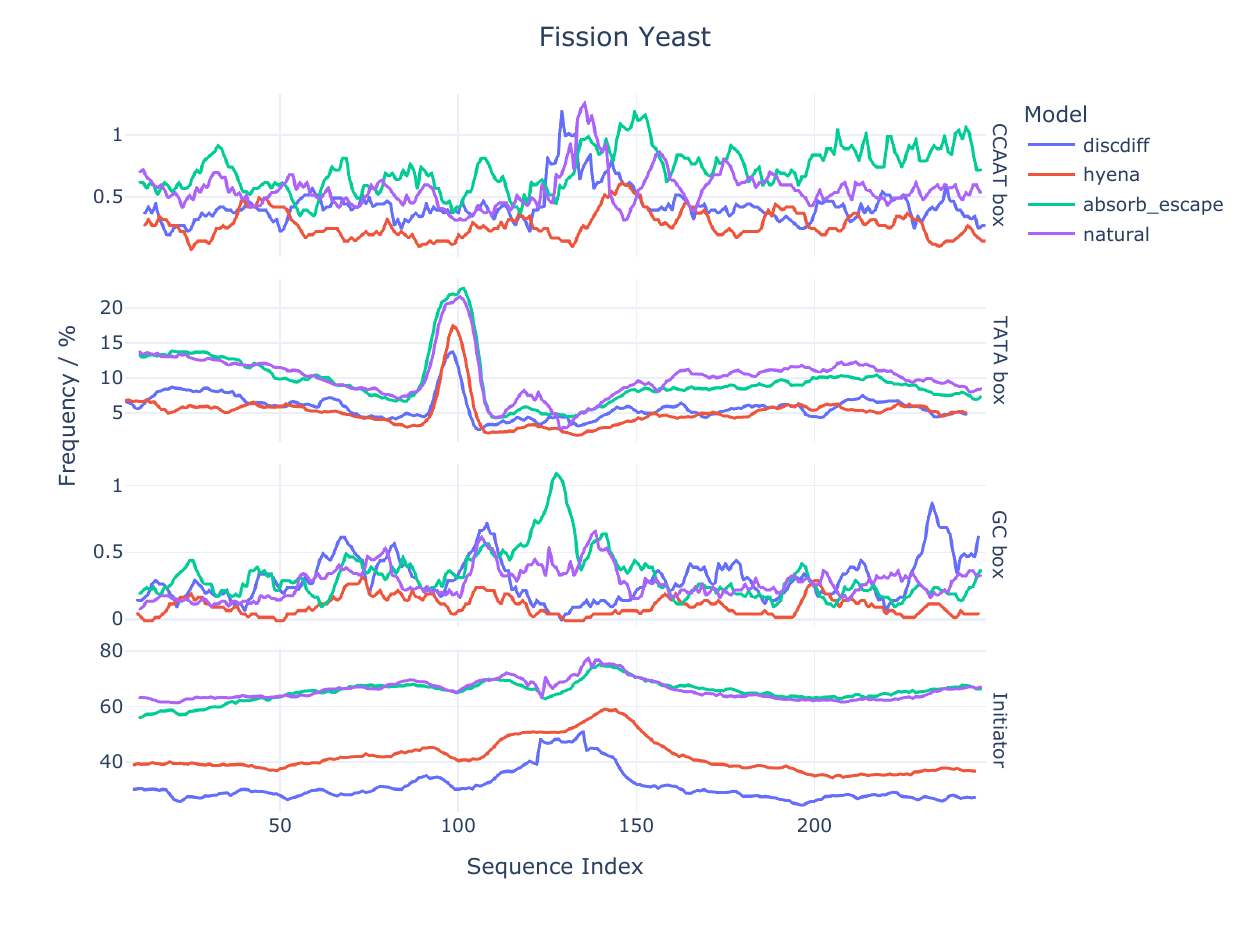}
    \caption{fission yeast}
    \label{fig:motif-distributions7}
\end{figure}

\begin{figure}[h]
    \centering
    \includegraphics[width=0.8\textwidth]{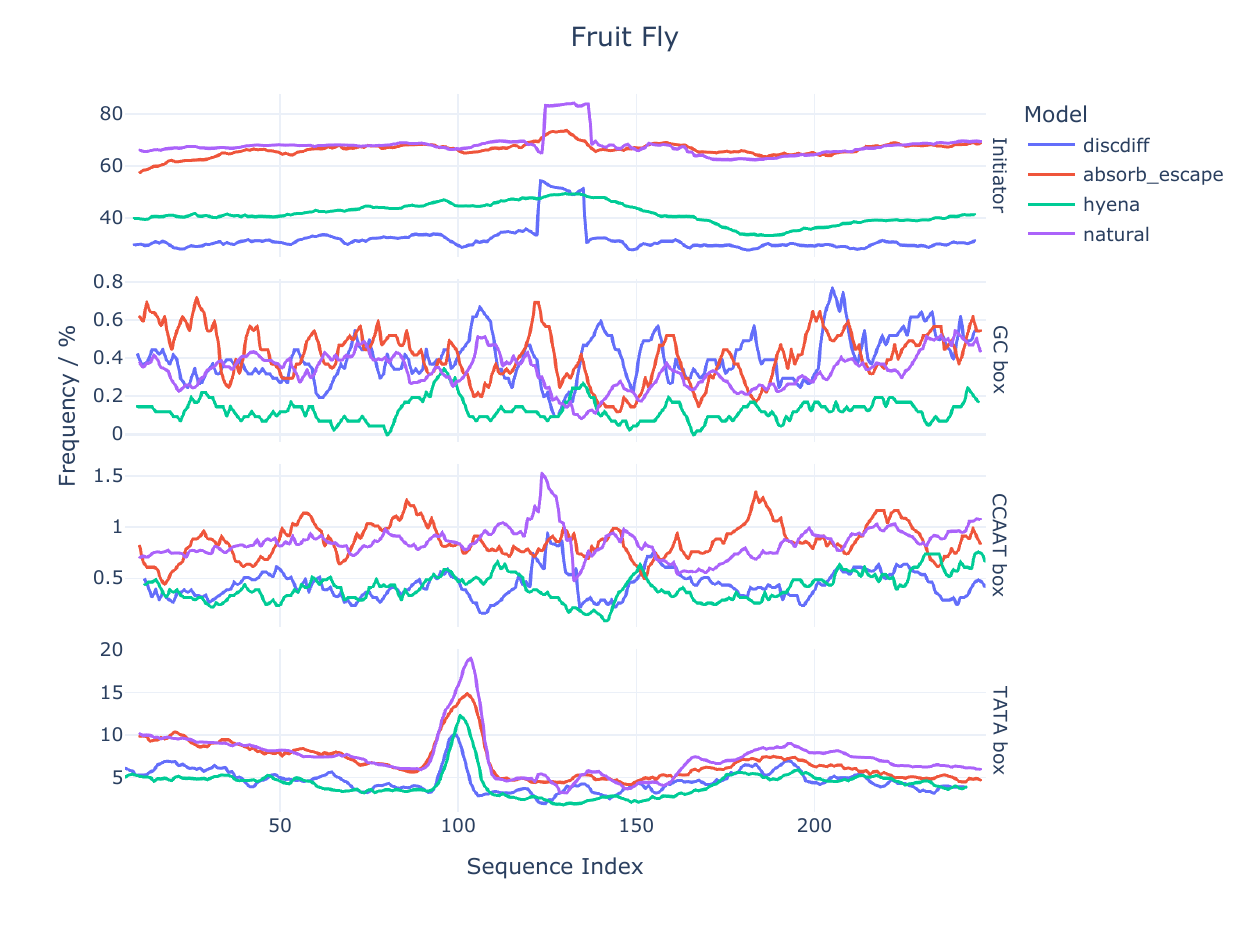}
    \caption{fruit fly}
    \label{fig:motif-distributions8}
\end{figure}

\begin{figure}[h]
    \centering
    \includegraphics[width=0.8\textwidth]{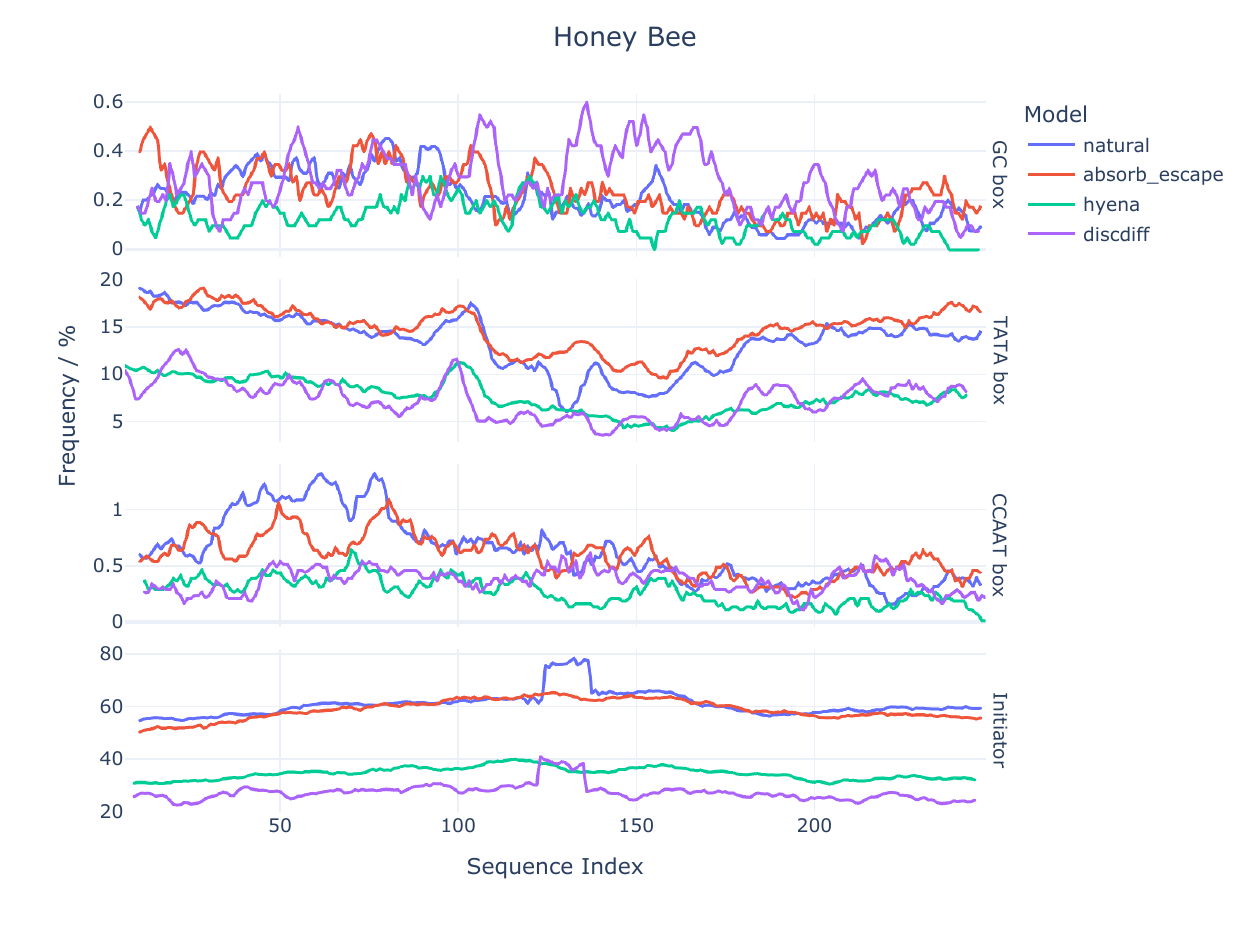}
    \caption{honey bee}
    \label{fig:motif-distributions9}
\end{figure}
\begin{figure}[h]
    \centering
    \includegraphics[width=0.8\textwidth]{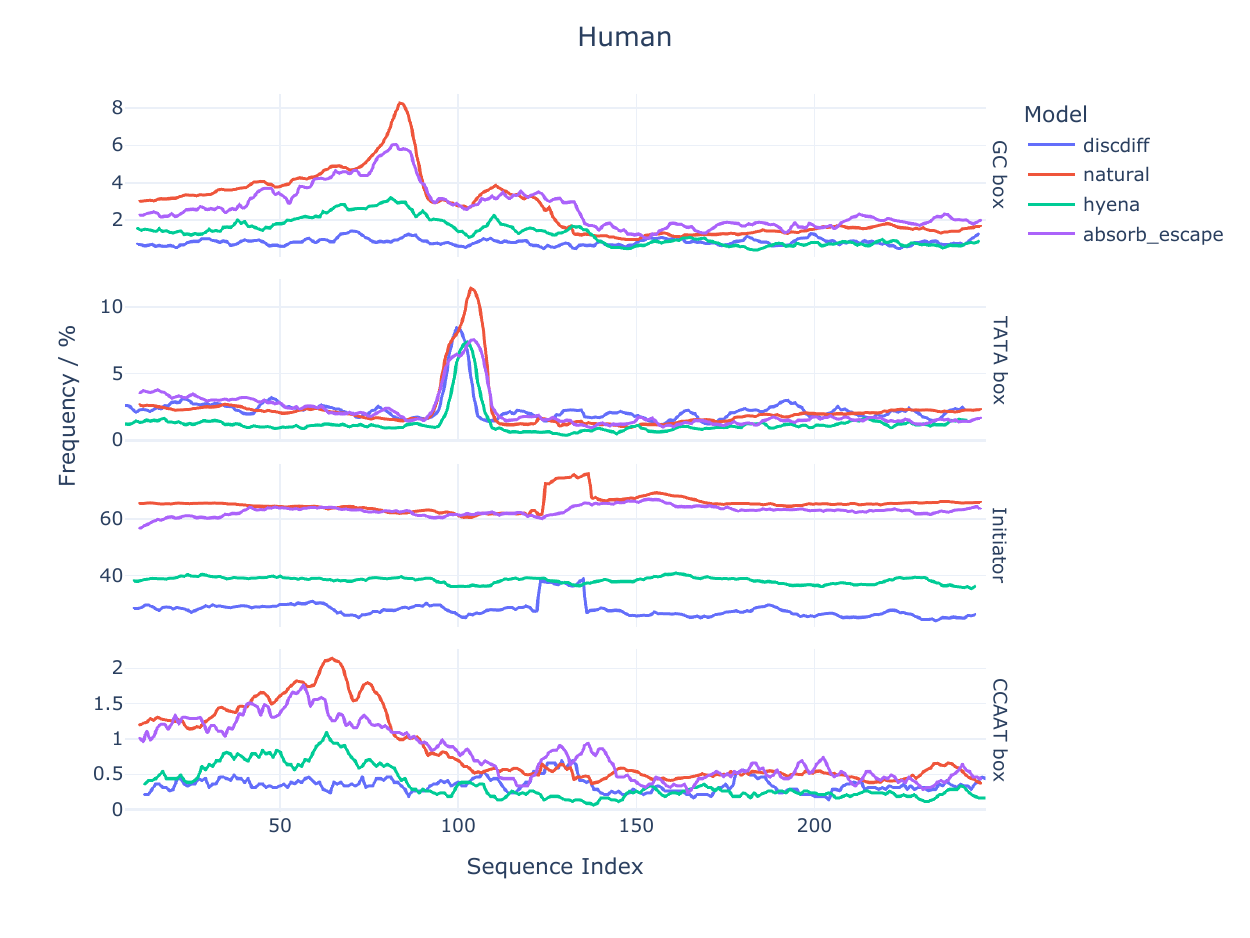}
    \caption{human}
    \label{fig:motif-distributions10}
\end{figure}
\begin{figure}[h]
    \centering
    \includegraphics[width=0.8\textwidth]{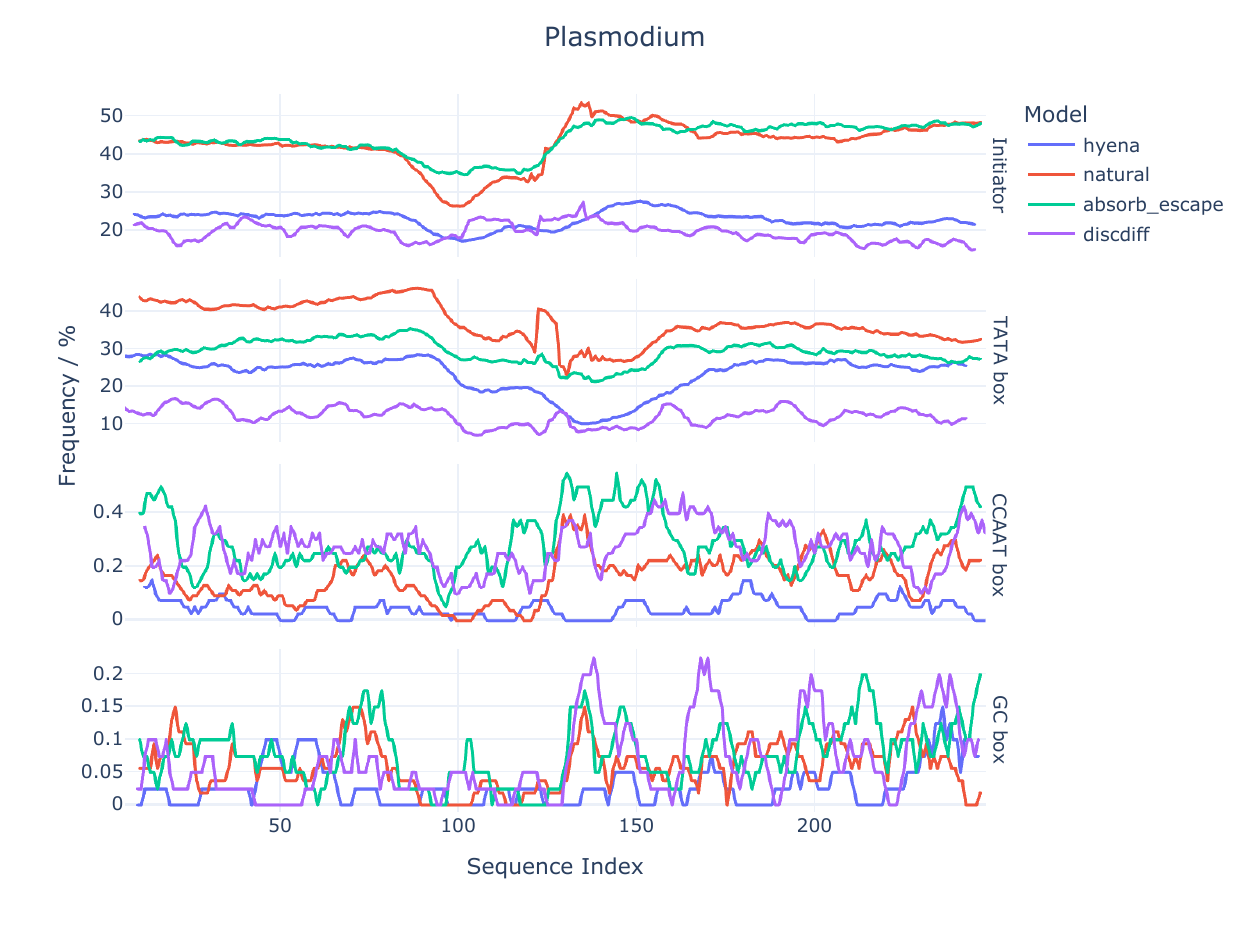}
    \caption{plasmodium}
    \label{fig:motif-distributions11}
\end{figure}
\begin{figure}[h]
    \centering
    \includegraphics[width=0.8\textwidth]{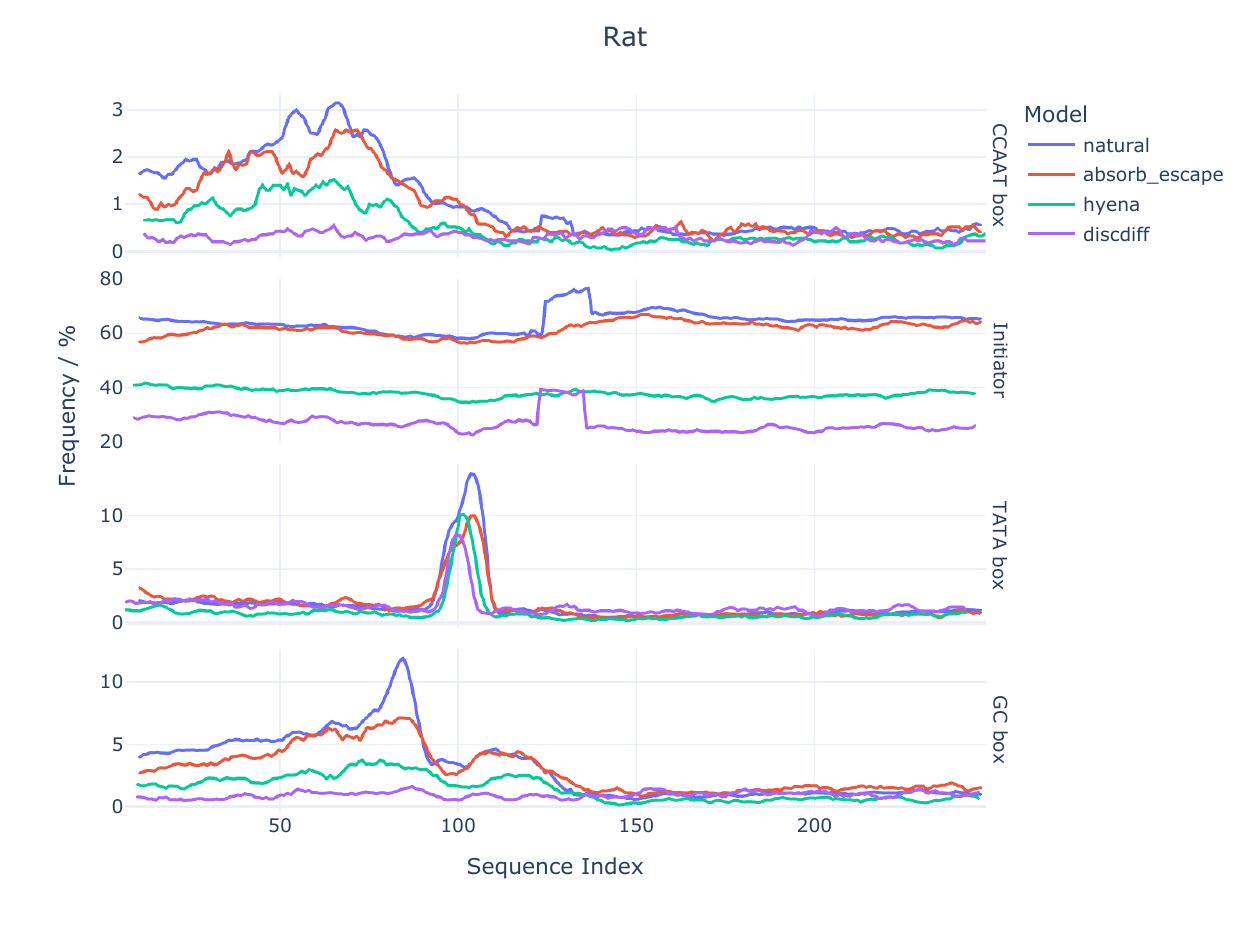}
    \caption{rat}
    \label{fig:motif-distributions12}
\end{figure}
\begin{figure}[h]
    \centering
    \includegraphics[width=0.8\textwidth]{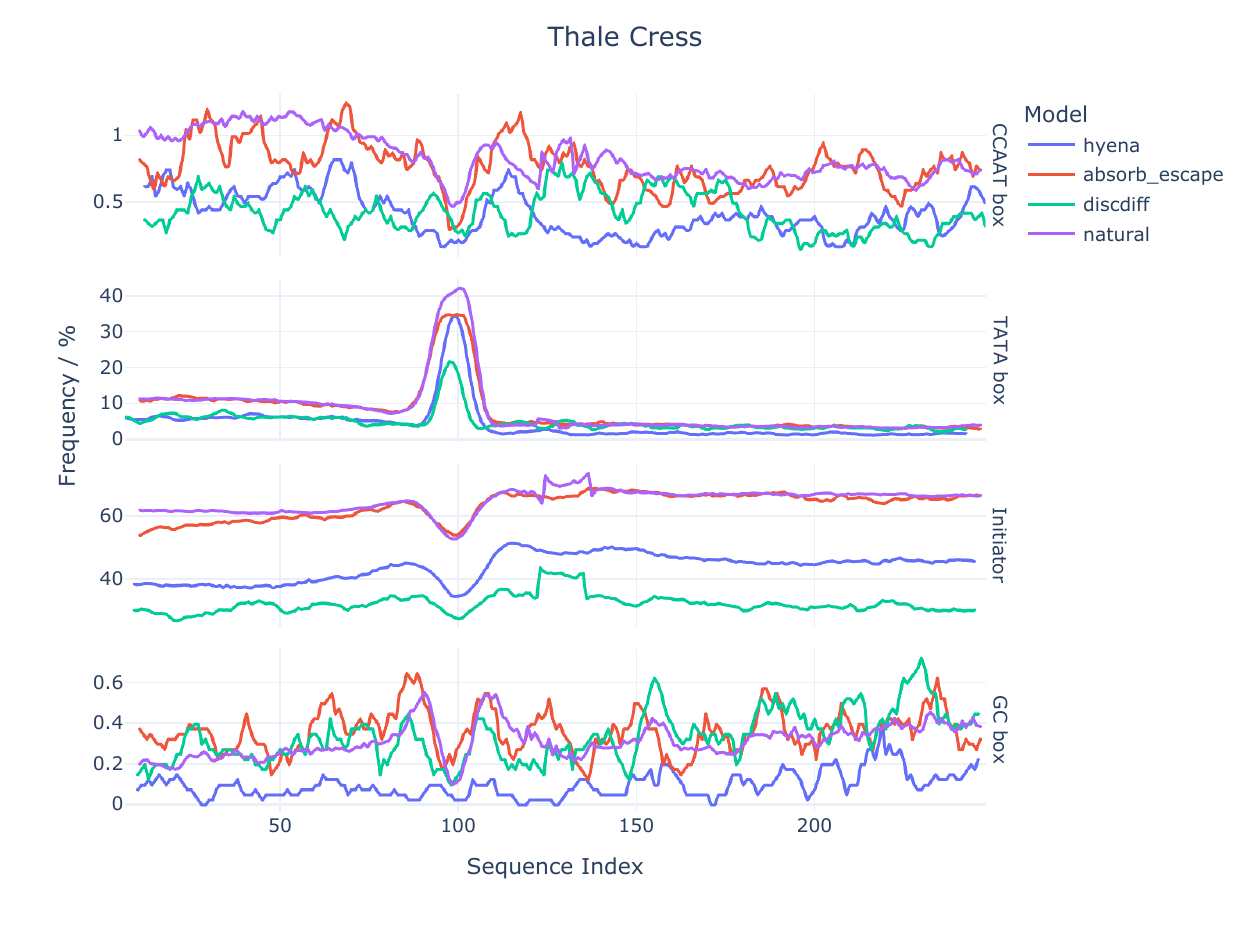}
    \caption{thale cress}
    \label{fig:motif-distributions13}
\end{figure}
\begin{figure}[h]
    \centering
    \includegraphics[width=0.8\textwidth]{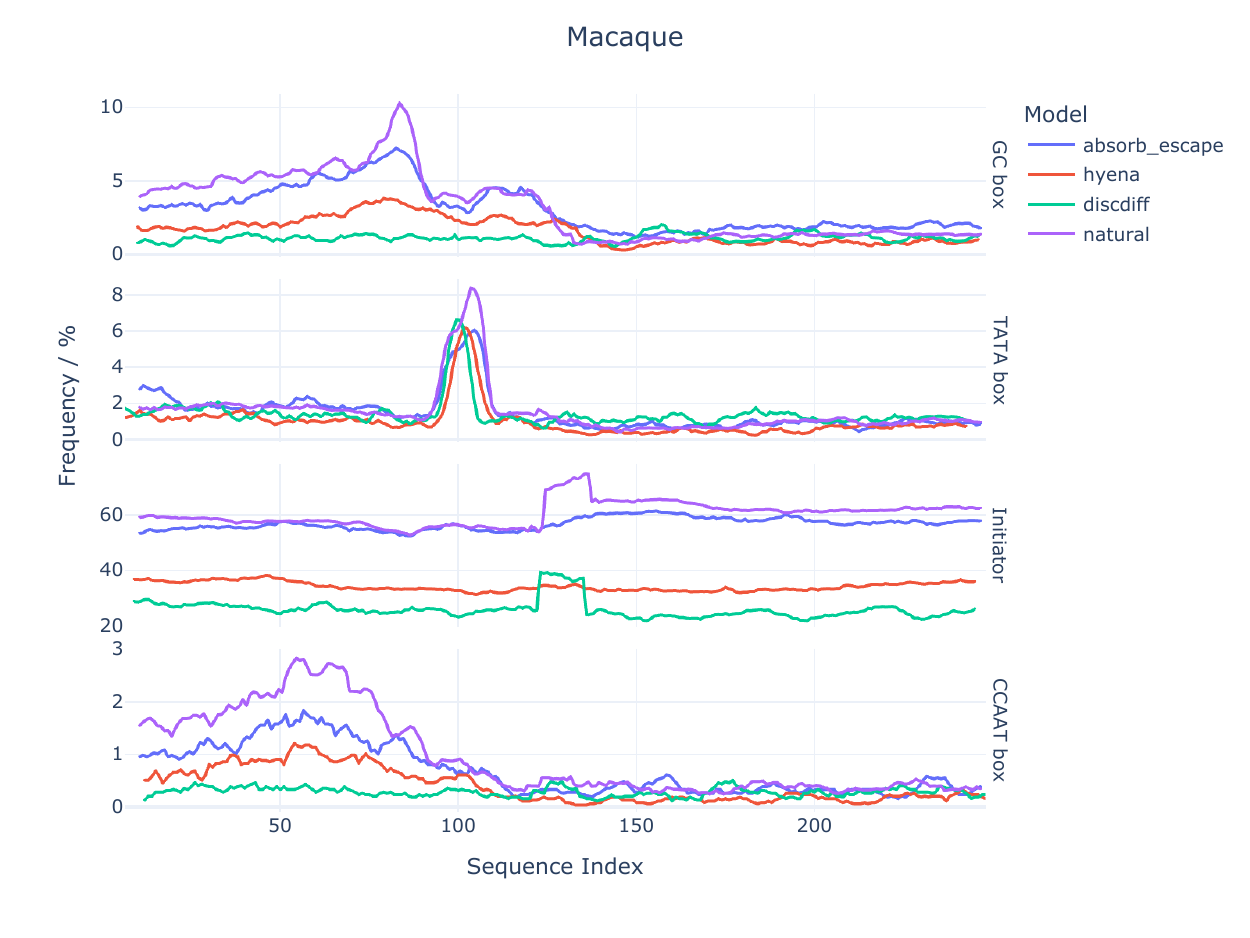}
    \caption{macaque}
    \label{fig:motif-distributions14}
\end{figure}

\begin{figure}[h]
    \centering
    \includegraphics[width=0.8\textwidth]{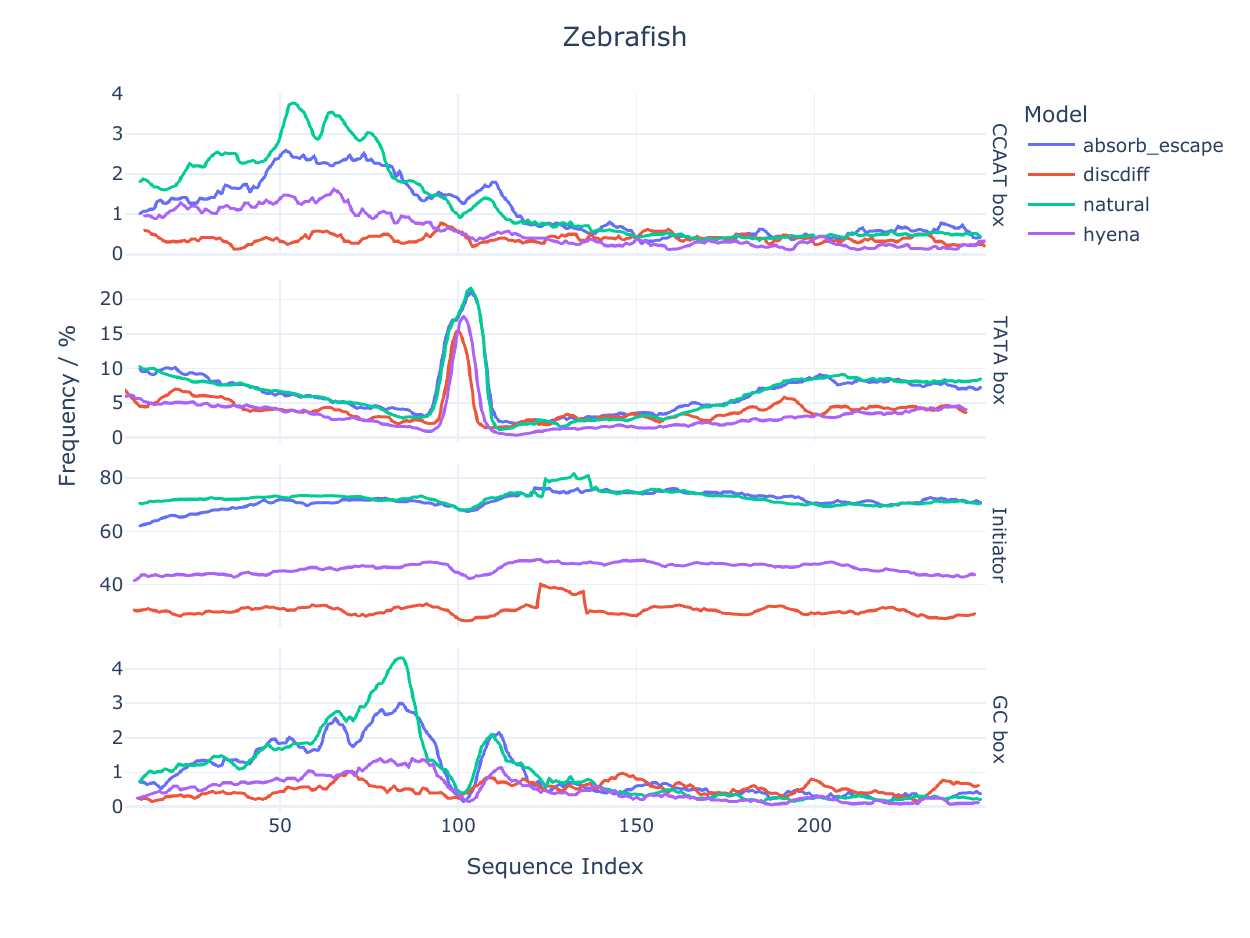}
    \caption{ zebrafish}
    \label{fig:motif-distributions15}
\end{figure}


\end{document}